\documentclass[traditabstract]{aa}
\usepackage{natbib}
\usepackage{epsfig}
\usepackage{txfonts}
\usepackage{graphicx}
\usepackage{float}

\usepackage{color,hyperref}
\definecolor{darkblue}{rgb}{0.0,0.0,0.4}
\hypersetup{colorlinks,breaklinks, linkcolor=darkblue,urlcolor=darkblue, anchorcolor=darkblue,citecolor=darkblue}

\def\ms{\,m\,s$^{-1}$}         
\def\kms{\,km\,s$^{-1}$}         
\def\ms{\hbox{\,m\,s$^{-1}\,$}}         
\def\cms{\hbox{\,cm\,s$^{-1}$}}       
\def\m2s2{\hbox{\,m$^{2}$\,s$^{-2}$}} 
\def\kms{\hbox{\,km\,s$^{-1}$}}       
\def\vsini{\hbox{$\upsilon \sin i_{\star}\;$}}      
\def\Msun{\hbox{$\mathrm{M}_{\odot}~$}}             
\def\Rsun{\hbox{$\mathrm{R}_{\odot}~$}}
\def\Mjup{\hbox{$\mathrm{M}_{\rm Jup}$~}}
\def\Rjup{\hbox{$\mathrm{R}_{\rm Jup}$~}}
\def\Rearth{\hbox{$\mathrm{R}_{\oplus}$~}}

\def\teff{T$_{\rm eff}~$}
\def\logg{$\log g$}
\def\met{[Fe/H]~}

\defcitealias{2011ApJ...738..170M}{M\&J11}
\defcitealias{2012arXiv1202.5852B}{Ba12}

\bibpunct{(}{)}{;}{a}{}{,}
\begin{document}
\title{\textsc{SOPHIE} velocimetry of \textit{Kepler} transit candidates \thanks{Based on observations made with SOPHIE on the 1.93-m telescope at Observatoire de Haute-Provence (CNRS), France}}
\subtitle{VII. A false-positive rate of 35\% for \textit{Kepler} close-in giant exoplanet candidates}

\author{
 A. Santerne \inst{1,3} 
\and R.~F. D\'iaz \inst{1, 2, 3}
\and C. Moutou \inst{1}
\and F. Bouchy \inst{2,3}
\and G. H\'ebrard \inst{2,3}
\and J.-M. Almenara \inst{1}
\and A.~S. Bonomo \inst{1}
\and M. Deleuil \inst{1}
\and N.~C. SantosÊ\inst{4, 5}}
\institute{
Laboratoire d'Astrophysique de Marseille, Universit\'e d'Aix-Marseille \& CNRS, UMR7326, 38 rue F. Joliot-Curie, 13388 Marseille Cedex 13, France 
\and Institut d'Astrophysique de Paris, UMR7095 CNRS, Universit\'e Pierre \& Marie Curie, 98bis boulevard Arago, 75014 Paris, France
\and Observatoire de Haute-Provence, Universit\'e d'Aix-Marseille \& CNRS, 04870 Saint Michel l'Observatoire, France
\and Centro de Astrof\'isica, Universidade do Porto, Rua das Estrelas, 4150-762 Porto, Portugal
\and Departamento de F\'isica e Astronomia, Faculdade de Ci\^encias, Universidade do Porto, Portugal}
\date{Received: 15 May 2012; Accepted: 26 June 2012}

\offprints{Alexandre~Santerne\\
 \email{alexandre.santerne@oamp.fr}}

\abstract{The false-positive probability (FPP) of \textit{Kepler} transiting candidates is a key value for statistical studies of candidate properties. A previous investigation of the stellar population in the \textit{Kepler} field has provided an estimate for the FPP of less than 5\% for most of the candidates. We report here the results of our radial velocity observations on a sample of 46 \textit{Kepler} candidates with a transit depth greater than 0.4\%, orbital period less than 25 days and host star brighter than \textit{Kepler} magnitude 14.7. We used the SOPHIE spectrograph mounted on the 1.93-m telescope at the Observatoire de Haute-Provence to establish the nature of the transiting candidates. In this sample, we found five undiluted eclipsing binaries, two brown dwarfs, six diluted eclipsing binaries, and nine new transiting planets that complement the 11 already published planets. The remaining 13 candidates were not followed-up or remain unsolved due to photon noise limitation or lack of observations. From these results we computed the FPP for \textit{Kepler} close-in giant candidates to be $34.8\%\pm6.5\%$. We aimed to investigate the variation of the FPP for giant candidates with the longer orbital periods and found that it should be constant for orbital periods between 10 and 200 days. This significant disagrees with the previous estimates. We discuss the reasons for this discrepancy and the possible extension of this work toward smaller planet candidates. Finally, taking the false-positive rate into account, we refined the occurrence rate of hot jupiters from the \textit{Kepler} data.

}

\titlerunning{FPP of \textit{Kepler} giant planet candidates}
\authorrunning{A.~Santerne et al.}

\maketitle

\section{Introduction}
\label{intro}
Since 2009, the \textit{Kepler} space mission \citep{2010Sci...327..977B} is monitoring more than 150,000 stars with high-precision photometry to search for transiting earth-like planets in the habitable zone. The unprecedented photometric precision reached by \textit{Kepler} has permitted the discovery of the first validated transiting planet in the habitable zone \citep{2012ApJ...745..120B} out of the solar system as well as extrasolar planets with radii of about the size of the Earth \citep{2012Natur.482..195F, 2012ApJ...747..144M}. The large number of planet candidates discovered so far \citep[2321, ][hereafter \citetalias{2012arXiv1202.5852B}]{2012arXiv1202.5852B} was used to estimate the occurrence of planets around solar-type stars \citep[e.g.][from the list of 1235 KOIs]{2011arXiv1103.2541H}, in relative agreement with Doppler surveys \citep{2010Sci...330..653H, 2011arXiv1109.2497M}. However, this statistical analysis assumes that most planet candidates are real planets or at least that the impostor rate is negligible. The experience gained from ground-based surveys or the pioneer space mission \textit{CoRoT} \citep{2006cosp...36.3749B} show that these false positives are mainly caused by undiluted eclipsing binaries with a low-mass stellar companion \citep{2008A&A...488L..47M, 2009A&A...506..337A} or diluted eclipsing binaries (so-called ``blends''), whose eclipse depth is diluted with the target flux and can therefore mimic a planetary transit \citep[e.g.][]{2011A&A...534A..67T}. Statistical analysis of stellar populations in the Milky Way can provide estimates of the false-positive rate \citep{2003ApJ...593L.125B}. A statistical study of the false-positive probability (hereafter FPP) of the \textit{Kepler} candidates has been performed by \citet{2011ApJ...738..170M} (hereafter \citetalias{2011ApJ...738..170M}). They found an overall FPP of less than 10\% for 90\% of the \textit{Kepler} candidates with a median value close to 5\%.\\

To establish the planetary nature of a transiting candidate, one must measure its mass through radial velocity (RV) follow-up \citep{2009IAUS..253..129B} or by using the transit-time variation technique \citep{2010Sci...330...51H}. Another solution is to ``validate'' the candidate by excluding all false-positive scenarios with a significant confidence level \citep[the so-called ``\texttt{BLENDER}'' technique, e.g.][]{2011ApJ...727...24T, 2011ApJS..197....5F}. Establishing or validating the nature of planetary candidates for a substantial sample of \textit{Kepler} candidates can improve the true fraction of \textit{Kepler} false positives and can thus improve the interpretation of \textit{Kepler} planet population.\\

In this paper, we first present our selection of \textit{Kepler} giant planet candidates (section \ref{keplerselect}) and their nature (section \ref{sophieobs}). We establish the candidate's nature using the SOPHIE spectrograph at the Observatoire de Haute-Provence. Our results allow us to independently measure the \textit{Kepler} FPP of these giant candidates (section \ref{section:FPP}) and to compare it with other estimates (\ref{MJ11}, \ref{otherFPP}). We finally estimate the trend of the false-positive rate for longer-period giant planets (section \ref{longPeriod}) and the impact of this new FPP value on the exoplanet statistics (section \ref{exoStat}), especially the occurrence rate (section \ref{sect:occurrence})

\section{Selection of \textit{Kepler} candidates}
\label{keplerselect}

Initially, a first list of 306 \textit{Kepler} planetary candidates was published in June 2010 by \citet{2011ApJ...728..117B}. This list contained only candidate transiting stars with \textit{Kepler} magnitude K$_{p} > 14$. Out of this list, our team selected four candidates for follow-up with the SOPHIE spectrograph during the summer 2010 \citep{bouchy2011}. These observations led to the discovery of the two first planets established from the public data: KOI-428b \citep{2011A&A...528A..63S} and KOI-423b \citep{bouchy2011}. In February 2011, the public list of candidates was extended to 1235 candidates \citep{2011ApJ...736...19B}, and to 2321 candidates in February 2012 \citepalias{2012arXiv1202.5852B}, including the brighter targets.\\

From the  \citet{2011ApJ...736...19B} list and then from the \citetalias{2012arXiv1202.5852B} list of Kepler Objects of Interest (KOI), we defined a sample of candidates for follow-up with the SOPHIE spectrograph: we first removed all targets with K$_{p} > 14.7$, which corresponds to the magnitude limit of the SOPHIE spectrograph \citep{2011EPJWC..1102001S}. About 49\% of \textit{Kepler} candidates are orbiting around such faint stars. We then rejected all candidates whose transit depth is shallower than 0.4\% in flux. We preferred selecting targets based on their transit depth rather than their expected radius because candidate radii are derived using the estimate of the stellar radius which can have up to 30\% uncertainty \citep{2011ApJ...736...19B}. We decided first of all to focus on short-orbital-period giant planet candidates and kept only candidates with orbital period shorter than 25 days. Indeed, all known transiting planets with a transit deeper than 0.4\% and an orbital period of less than 25 days have a radial velocity semi-amplitude greater than 10\ms. This limit is close to the photon noise reached by SOPHIE in 1-h exposure time on a $\sim 13^{th}$ magnitude star. Out of the 2321 KOIs, only $\sim 3.8\%$ present both a star brighter than $K_{p}=14.7$ and a transit depth greater than 0.4\%. If we had kept only the shortest orbital period ($P < 25$ days), $\sim 2.3\%$ candidates would have remained. Finally, we removed the eight candidates with a vetting flag\footnote{According to \citet{2011ApJ...736...19B}, a vetting flag of 4 means : ``Insufficient follow-up to perform full suite of vetting tests''.} of 4 in \citet{2011ApJ...736...19B} that match all the previous criteria. Indeed, of these eight lowest priority candidates, most are either clearly eclipsing binaries with transit depths of up to 8\% or show a high level of variability in their light curve that is caused by a fast rotating or pulsating host star.\\

Only 46 candidates fullfil all criteria, which corresponds to about $2\%$ of the total list of the 2321 candidates as of February 2012 and to about 22\% of all giant planet candidates (with depth $> 0.4\%$) found by \textit{Kepler} up to now. About 60\% of the \textit{Kepler} giant planet candidates are orbiting stars fainter than $Kp=14.7$. The 46 selected candidates, with their parameters, are listed in Table \ref{Saturn}. We note that only two candidates in this sample are in a multiple system: KOI-94.01 and KOI-377.01 (Kepler-9b). We also note that only two candidates were added with the updated list of KOIs from \citetalias{2012arXiv1202.5852B}: KOI-554.01 and KOI-1786.01.

\begin{table*}[h]
\centering
\renewcommand{\footnoterule}{}                          
\begin{minipage}[c]{2\columnwidth} 
\caption{List of the 46 selected candidates for follow-up with SOPHIE with K$_{p} < 14.7$ \& depth$\ > 0.4$ \% \& $P < 25$ d \& Vetting flag $\neq 4$.}
\centering
\begin{tabular}{ccccccccccc}
\hline
\hline
KIC  &  KOI  &  Period$^{\ast}$  &  K$_{p}^{\ast}$ & Depth$^{\ast}$ & R$_{p}^{\ast}$ & V flag$^{\ast\ast}$ & FPP$^{\star}$ & Object nature$^{\dag}$ & Mass & Reference$^{\ddag}$ \\
  &   &  [days]  &   &  [\%] & [\Rjup] & & [\%] &  & [\Mjup] &\\
\hline
11446443 & 1.01 & 2.47 & 11.3 & 1.42 & 1.28 & 1 & 1.0 & planet / TrES-2 & 1.20 & OD06 \\
10666592 & 2.01 & 2.20 & 10.5 & 0.67 & 1.99 & 1 & 0.5 & planet / HAT-P-7b & 1.80 & P\'a08\\
10748390 & 3.01 & 4.89 & 9.1 & 0.42 & 0.42 & 1 & 6.0 & planet / HAT-P-11b & 0.08 & Ba10\\
6922244 & 10.01 & 3.52 & 13.6 & 0.94 & 1.42 & 1 & 0.9 & planet / Kepler-8b & 0.60 & Je10\\
5812701 & 12.01 & 17.86 & 11.4 & 0.93 & 1.18 & 3 & 4.5 & no var & $<$ 26.7 & this work\\
9941662 & 13.01 & 1.764 & 10.0 & 0.46 & 2.03 & 2 & 0.5 & planet / KOI-13b & $<$9.2 & Ma11, Sh11, Mi12\\
10874614 & 17.01 & 3.23 & 13.0 & 1.07 & 0.99 & 2 & 0.8 & planet / Kepler-6b & 0.67 & Du10\\
8191672 & 18.01 & 3.55 & 13.4 & 0.72 & 1.55 & 1 & 1.0 & planet / Kepler-5b & 2.11 & Ko10\\
11804465 & 20.01 & 4.44 & 13.4 & 1.67 & 1.56 & 2 & 1.1 & planet / Kepler-12b & 0.43 & Fo11\\
9631995 & 22.01 & 7.89 & 13.4 & 1.06 & 1.00 & 2 & 2.8 & unknown / no FUp & -- &\\
6056992 & 51.01 & 10.43 & 13.8 & 2.58 & 1.87 & 3 & 2.7 & blend & -- & this work\\
11554435 & 63.01 & 9.43 & 11.6 & 0.41 & 0.56 & 2 & 3.2 & unknown / No FUp & -- & \\
6462863 & 94.01 & 22.34 & 12.2 & 0.57 & 0.83 & 2 & 4.8 & unknown / no FUp & -- & \\
5780885 & 97.01 & 4.89 & 12.9 & 0.74 & 1.43 & 1 & 0.9 &  planet / Kepler-7b & 0.43 &La10\\
8359498 & 127.01 & 3.58 & 13.9 & 1.16 & 0.97 & 2 & 0.9 & unknown / no FUp & -- &\\
11359879 & 128.01 & 4.94 & 13.8 & 1.12 & 1.07 & 2 & 1.1 & planet / Kepler-15b & 0.66 & En11\\
7778437 & 131.01 & 5.01 & 13.8 & 0.69 & 0.86 & 3 & 1.3 & no var & $<$ 14.3 & this work\\
9818381 & 135.01 & 3.02 & 14.0 & 0.79 & 0.94 & 2 & 0.8  & planet / KOI-135b & 3.23 &Bon12\\
9651668 & 183.01 & 2.68 & 14.3 & 1.83 & 1.03 & 2 & 0.9 & unknown / no FUp & -- &\\
5771719 & 190.01 & 12.27 & 14.1 & 1.15 & 1.39 & 2 & 5.5 & blend & -- & this work\\
7950644 & 192.01 & 10.29 & 14.2 & 1.00 & 0.84 & 2 & 4.5 & no var & $<$ 0.6 & this work\\
9410930 & 196.01 & 1.86 & 14.5 & 1.08 & 0.88 & 2 & 0.7 & planet / KOI-196b & 0.55 & Sa11 \\
2987027 & 197.01 & 17.28 & 14.0 & 1.08 & 0.70 & 2 & 4.6 & no var & $<$ 0.32 & this work\\
6046540 & 200.01 & 7.34 & 14.4 & 0.85 & 0.79 & 2 & 2.4 & planet / KOI-200b & 0.44 & H\'e+ \\
6849046 & 201.01 & 4.23 & 14.0 & 0.60 & 0.89 & 3 & 0.9 & no var & $<$ 0.6 & this work\\
7877496 & 202.01 & 1.72 & 14.3 & 1.03 & 1.02 & 2 & 0.6 & planet / KOI-202b & 0.88 & H\'e+\\
10619192 & 203.01 & 1.49 & 14.1 & 2.09 & 1.32 & 2 & 0.7 & planet / Kepler-17b & 2.47 & D\'e11, Bon12\\
9305831 & 204.01 & 3.25 & 14.7 & 0.72 & 0.65 & 2 & 1.1 & planet / KOI-204b & 1.02 & Bon12\\
7046804 & 205.01 & 11.72 & 14.5 & 1.00 & 0.67 & 2 & 5.3 & BD & $\sim$ 35 & D\'i+\\
5728139 & 206.01 & 5.33 & 14.5 & 0.50 & 0.70 & 2 & 1.3 & planet / KOI-206b & 2.9 & H\'e+\\
11046458 & 214.01 & 3.31 & 14.2 & 0.58 & 0.64 & 2 & 1.0 & unknown / NoFUp & -- &\\
10616571 & 340.01 & 23.67 & 13.1 & 2.12 & 1.50 & 3 & 100 & SB1 & 560 & this work\\
3323887 & 377.01 & 19.26 & 13.8 & 0.75 & 0.74 & 1 & 4.5 & planet / Kepler-9b & 0.25 & Ho10\\
5449777 & 410.01 & 7.22 & 14.5 & 0.41 & 3.57 & 2 & 1.9 & no var & $<$ 3.4 & Bou11\\
7975727 & 418.01 & 22.42 & 14.5 & 1.22 & 0.84 & 2 & 6.8 & blend & -- & this work\\
8219673 & 419.01 & 20.13 & 14.5 & 0.77 & 2.77 & 2 & 4.4 &  SB1 & 723 & this work\\
9478990 & 423.01 & 21.09 & 14.3 & 0.91 & 0.84 & 2 & 5.9 & planet / KOI-423b & 18.00 & Bou11\\
9967884 & 425.01 & 5.43 & 14.7 & 1.23 & 1.05 & 2 & 1.7 & blend & -- & this work\\
5443837 & 554.01 & 3.66 & 14.5 & 0.54 & 0.65 & -- & -- & BD & $\sim$ 80 & D\'i+\\
5441980 & 607.01 & 5.89 & 14.4 & 0.66 & 0.57 & 3 & 1.8 & SB1 & 120 & this work\\
5608566 & 609.01 & 4.40 & 14.5 & 0.43 & 1.32 & 3 & 1.3 & blend & -- & this work\\
6309763 & 611.01 & 3.25 & 14.0 & 0.43 & 1.01 & 2 & 1.1 & no var & $<$1.5 & this work\\
6752502 & 667.01 & 4.31 & 13.8 & 1.01 & 0.47 & 3 & 10.2 & blend & -- & this work\\
7529266 & 680.01 & 8.60 & 13.6 & 0.44 & 0.64 & 2 & 3.0 & planet / KOI-680b & 0.62 & H\'e+\\
8891278 & 698.01 & 12.72 & 13.8 & 0.78 & 1.02 & 2 & 4.6 & SB1 & 859 & this work\\
3128793 & 1786.01 & 24.68 & 14.6 & 0.82 & 0.39 & -- & -- & SB1 & 244 & this work\\
\hline
\hline
\end{tabular}
\vspace{-0.3cm}
\footnotetext{$^{\ast}$ Orbital period, Kepler magnitude (K$_{p}$), transit depth and expected planetary radius from \citet{2011ApJ...736...19B} and \citetalias{2012arXiv1202.5852B}. The expected planetary radius may have uncertainty up to 30\% \citep{2011ApJ...736...19B}.}
\footnotetext{$^{\ast\ast}$ Vetting flag from \citet{2011ApJ...736...19B}: '1' for ``confirmed and published planet''; '2'	 for ``Strong probability candidate, cleanly passes tests that were applied''; '3' for ``moderate probability candidate, not all tests cleanly passed but no definite test failures''.}
\footnotetext{$^{\star}$ False Positive Probability as estimated by \citet{2011ApJ...738..170M}.}
\footnotetext{$^{\dag}$ no var: no significant RV variation; blend: triple system or background eclipsing binary; SB1: single-line spectroscopic binary; no FUp: no follow-up observation with SOPHIE; BD: Brown dwarf with mass in between 25\Mjup and 80\Mjup.}
\footnotetext{$^{\ddag}$Ba10: \citet{2010ApJ...710.1724B}; Bon12: \citet{bonomo2012}; Bor11: \citet{2011ApJ...736...19B}; Bou11: \citet{bouchy2011}; D\'e11: \citet{2011ApJS..197...14D}; D\'i+: D\'iaz et al. (in prep.); Du10: \citet{2010ApJ...713L.136D}; En11: \citet{2011ApJS..197...13E}; Fo11: \citet{2011ApJS..197....9F}; H\'e+: H\'ebrard et al. (in prep.); Ho10: \citet{2010Sci...330...51H}; Je10: \citet{2010ApJ...724.1108J}; Ko10: \citet{2010ApJ...713L.131K}; La10: \citet{2010ApJ...713L.140L}; Ma11: \citet{2011arXiv1110.3512M}; Mi12: \citet{2012arXiv1202.1760M}; OD06: \citet{2006ApJ...651L..61O}; P\'a08: \citet{2008ApJ...680.1450P}; Sa11: \citet{Santerne2011}; Sh11: \citet{2011AJ....142..195S}.}
\label{Saturn}      
\end{minipage}
\end{table*}

\section{SOPHIE observations}
\label{sophieobs}

\subsection{Observations and data reduction}

We started a new large program in early 2011 to perform spectroscopic follow-up observations on the 46 selected \textit{Kepler} targets with the SOPHIE spectrograph \citep{2008SPIE.7014E..17P, 2009A&A...505..853B} mounted on the 1.93-m telescope at the Observatoire de Haute-Provence, France. Observations were conducted from 2011, February 24 to 2012, May 2 using the high-efficiency mode ($R \sim 39~000$ at 550nm) of SOPHIE\footnote{prog. IDs: 11A.PNP.MOUT, 11B.PNP.MOUT, 12A.PNP.MOUT}. Spectra were reduced with the online standard pipeline and radial velocities were obtained by computing the weighted cross-correlation function (CCF) of the spectra with a numerical spectral mask of a G2V star \citep{1996A&AS..119..373B, 2002A&A...388..632P}. For some candidates, we also correlated the CCF using a F0V and K5V mask to test the mask effect \citep{2009IAUS..253..129B}.\\

Several spectra were significantly affected by the scattered moon light and corrected using the same technique as in \citet{2011A&A...528A..63S} and \citet{2010arXiv1006.2949B}. We tried to keep the signal-to-noise ratio (S/N) as constant as possible to limit the charge transfer inefficiency effect of the CCD camera \citep{2009EAS....37..247B}. We corrected the radial velocities of a given target that were computed from different S/N spectra with equation \ref{CTIfit}. This empirical function was calibrated with dedicated observations with SOPHIE at different S/N spectra of the daily blue sky:

\begin{equation}
\label{CTIfit}
\Delta_{RV}(\mathrm{S/N}_{550nm}) = -6.265\times(\mathrm{S/N}_{550nm})^{-1.71} [\kms],
\end{equation}
where $\mathrm{S/N}_{550nm}$ is the signal-to-noise ratio per pixel measured on the extracted spectrum in the range 10 -- 50.
We also computed the \vsini using Appendix B.1. in \citet{2010A&A...523A..88B}. The radial velocities are listed in Tables \ref{koi12} to \ref{koi1786}.\\

\subsection{Spectroscopic analysis}

For the candidates that do present a significant radial velocity variation, we performed a detailed spectroscopic analysis to determine their stellar parameters to derive the candidate parameters. For the low-rotating ones (namely KOI-192, KOI-197 and KOI-201) this analysis was based on iron line excitation and ionization equilibrium. This analysis made use of a grid of \citet{1997ESASP.419..193K} model atmospheres and of the 2002 version of the radiative transfer code MOOG \citep{1973ApJ...184..839S}. For details we refer the reader to \citet{2004A&A...415.1153S} and \citet{2008A&A...487..373S}.\\

The spectroscopic analysis is based on the measurement of line equivalent widths (EWs) for a list of selected \ion{Fe}{i} and \ion{Fe}{ii} lines. For this we employed the stacked SOPHIE spectra
used for the derivation of radial velocities. For KOI-197, the spectrum had a total S/N sufficiently high (around 50 at 6700\AA) for us to use the automatic code ARES \citep{2007A&A...469..783S} to measure the EWs for the almost 300 lines used in the analysis. For KOI-192 and KOI-201, given the low S/N (around 20) of the available spectra, we decided to adopt a more careful analysis. We adopted the shorter (but well-tested) line-list presented in \citet{2004A&A...415.1153S} and the EWs were carefully measured one by one using the splot tool of IRAF.\\

We finally determined the stellar mass, radius and age by comparing the \logg, \teff, and \met from the stellar analysis to the STAREVOL evolution tracks\,Ê\citep{2010ApJ...715.1539T}, in the (\teff, \logg) H-R diagram.\\

For the fast-rotating no-variation candidates (namely KOI-12, KOI-131 and KOI-611), the resulting co-added spectrum S/N was too low to allow spectral analysis. We therefore considered the stellar parameters from \citetalias{2012arXiv1202.5852B} in our analysis.

\section{Establishing the nature of KOIs}

In this section, we discuss the results obtained on individual candidates. Stellar masses and radii come from the \textit{Kepler} Input Catalog \citep{2011AJ....142..112B}, except for three mentioned above.

\subsection{Secure planets}

Our selection of 46 KOIs includes three pre-launch planets: KOI-1.01, KOI-2.01 and KOI-3.01, also named TrES-2 \citep{2006ApJ...651L..61O}, HAT-P-7b \citep{2008ApJ...680.1450P} and HAT-P-11b \citep{2010ApJ...710.1724B}, respectively. It also includes the four planets announced by the \textit{Kepler} team in early 2010 : KOI-18.01 / Kepler-5b \citep{2010ApJ...713L.131K}, KOI-17.01 / Kepler-6b \citep{2010ApJ...713L.136D}, KOI-97.01 / Kepler-7b \citep{2010ApJ...713L.140L} and KOI-10.01 / Kepler-8b \citep{2010ApJ...724.1108J}. One planet of the KOI-377 / Kepler-9 system \citep{2010Sci...330...51H} is also present in our sample. In 2011, the \textit{Kepler} team established the planetary nature of KOI-20.01 / Kepler-12b  \citep{2011ApJS..197....9F}, KOI-128.01 / Kepler-15b \citep{2011ApJS..197...13E} and KOI-203.01 / Kepler-17b \citep{2011ApJS..197...14D} that were also in our KOI selection. At the time of the publication of the latter one, eight SOPHIE spectra had been acquired that permit us to independently confirm and improve the planet and stellar parameters \citep{bonomo2012}.\\

Another planet from this list has also been characterized and announced: KOI-13.01, for which an upper-limit on the mass in the planet regime was performed using the photometric beaming effect \citep{2011AJ....142..195S, 2011arXiv1110.3512M} and ellipsoidal effect \citep{2012arXiv1202.1760M}.\\

Finally, based on our observations obtained with the SOPHIE spectrograph, we found nine new planets in this sample: KOI-135b \citep{bonomo2012}, KOI-196b \citep{Santerne2011}, KOI-200b (H\'ebrard et al., in prep.), KOI-202b (H\'ebrard et al., in prep.), KOI-203b / Kepler-17b \citep{bonomo2012}, KOI-204b \citep{bonomo2012}, KOI-206b (H\'ebrard et al., in prep.), KOI-423b \citep{bouchy2011} and KOI-680b (H\'ebrard et al., in prep.). We also found KOI-428b \citep{2011A&A...528A..63S}. Since its transit depth is about $0.3\%$ due to the large radius of the host star, this hot jupiter is not included in our sample.\\

For this study, we considered the upper mass limit of planet proposed by \citet{2011A&A...532A..79S} of up to 25\Mjup. Out of the 46 candidates we selected, 20 turned out to be planets.

\subsection{Brown dwarfs}

We found two interesting objects that have a mass in between 25\Mjup and 80\Mjup. We classified these candidates as new transiting brown dwarfs but did not consider them as planets for the computation of the false-positive rate. These objects are KOI-205.01 and KOI-554.01, which have a mass of about 35\Mjup and 80\Mjup, respectively (D\'iaz et al., in prep.).

\subsection{Undiluted binaries}

By following-up the \textit{Kepler} candidates, we found several targets that present strong radial velocity variations that are not compatible with a planetary scenario. These candidates are consequently undiluted eclipsing binaries:

\subsubsection{KOI-340.01}
KOI-340.01 is a candidate on an $\sim$23.7-day-period orbit with a depth of 2.12\%. We observed this target twice with SOPHIE (see Table \ref{koi340} and Fig. \ref{figSB}). We found a strong radial velocity variation with a semi-amplitude of $K = 34.577 \pm 0.074$\kms\, assuming a circular orbit. If the stellar mass is 1.15\Msun \citepalias{2012arXiv1202.5852B}, the transiting companion would have a mass of $0.70\pm0.04$\Msun, and thus be an eclipsing binary. No secondary peak is visible in the CCF nor a clear secondary eclipse in the \textit{Kepler} LC. We suspect that the host star is an evolved star which would also explain the quite long-transit duration (14.4h) and the shallow eclipse for this eclipsing binary. We note that this planetary candidate is also classified as an eclipsing binary in the \textit{Kepler} eclipsing binary catalog\footnote{\url{http://archive.stsci.edu/kepler/eclipsing_binaries.html}}. This candidate has been estimated by \citetalias{2011ApJ...738..170M} to be a diluted binary (see discussion section \ref{MJ11}) with a probability of 100\% due to the very low likelihood of the planet scenario.

\subsubsection{KOI-419.01}
We observed KOI-419 seven times with SOPHIE. KOI-419.01 orbits its host-star in an $\sim$20.1-day period. The radial velocities listed in Table \ref{koi419} and displayed in Fig. \ref{figSB} show a clear radial velocity variation in anti-phase with \textit{Kepler} ephemeris. We analyzed both the \textit{Kepler} PDC LC (Q1 to Q6) filtered using an iterative smoothing filter keeping the period of the signal, the SOPHIE RVs, and the available SDSS, 2MASS \citep{2006AJ....131.1163S} and WISE magnitudes. We modeled a binary scenario using stellar atmosphere models of \citet{2004astro.ph..5087C}, light-curve models from JKTEBOP \citep{2008MNRAS.386.1644S} and evolution stracks from \citet{2008A&A...482..883M} and \citet{2010ApJ...724.1030G}. The model parameters were fitted through a MCMC simulation (D\'iaz et al., in prep). We fixed the orbital period to the one published by \citetalias{2012arXiv1202.5852B} and fitted the eccentricity, the inclination, the argument of periastron, and the systemic radial velocity. We assumed a binary system with two stars with a similar metallicity, fixed to the solar value, and the same $\log(\mathrm{age})$ that we allowed to vary, and fitted the respective initial masses, system distance and reddening coefficient. We found that the data are compatible with an eccentric binary ($e \sim 0.33$) for which only the secondary eclipse occurs (see fig \ref{KOI-419}). All parameters derived from this combined analysis are provided in Table \ref{KOI-419params}. We found the primary and secondary mass to be $1.20\pm0.12$ \Msun and $0.70\pm0.07$ \Msun, respectively. The system is thus an eclipsing binary.

\subsubsection{KOI-607.01}
KOI-607.01 is an $\sim$5.9-day-period candidate. We took two SOPHIE measurements (see Table \ref{koi607} and Fig. \ref{figSB}) that present radial velocity variations in phase with \textit{Kepler} ephemeris. Assuming a circular orbit, we found $K = 13.45 \pm 0.06 \kms$. If the mass of the host star is 0.79 \Msun \citepalias{2012arXiv1202.5852B}, the mass of the companion is $0.106\pm0.006$\Msun. This interesting low-mass star does not show a significant secondary eclipse at phase 0.5.

\subsubsection{KOI-698.01}
We observed the candidate host KOI-698 three times with SOPHIE. The transits occur every $\sim$12.7 days. The measured radial velocities (see Table \ref{koi698} and Fig. \ref{figSB}) show a strong variation in anti-phase with the \textit{Kepler} ephemeris. As for KOI-419, we analyzed  the \textit{Kepler} PDC LC (Q2 to Q6) using the same filtering method, the SOPHIE RVs and magnitudes from SDSS, 2MASS \citep{2006AJ....131.1163S} and WISE. We performed the same simulation using the same fixed and free parameters. As for KOI-419, we found that the system is also compatible with an eccentric binary ($e\sim 0.34$) for which the primary transit is unseen (see fig. \ref{KOI-698}). All parameters derived from this combined analysis are provided in Table \ref{KOI-698params} and lead to a primary and secondary mass of $1.34\pm0.13$ \Msun and $1.14 \pm 0.1$ \Msun, respectively.\\

\subsubsection{KOI-1786.01}
\begin{figure}[h]
\begin{center}
\setlength{\tabcolsep}{0.0mm}
\begin{tabular}{cc}
\includegraphics[width=0.5\columnwidth]{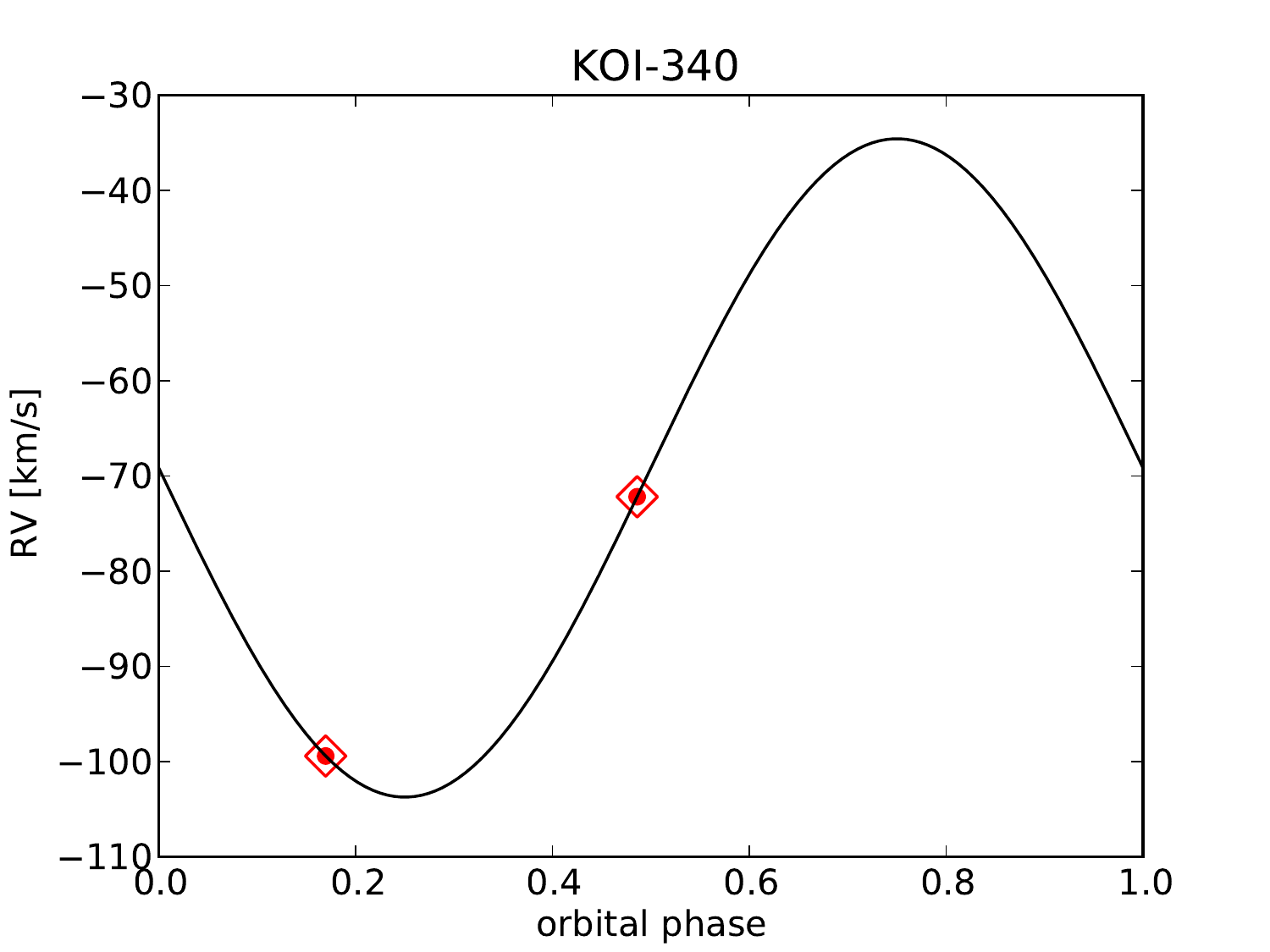} & \includegraphics[width=0.5\columnwidth]{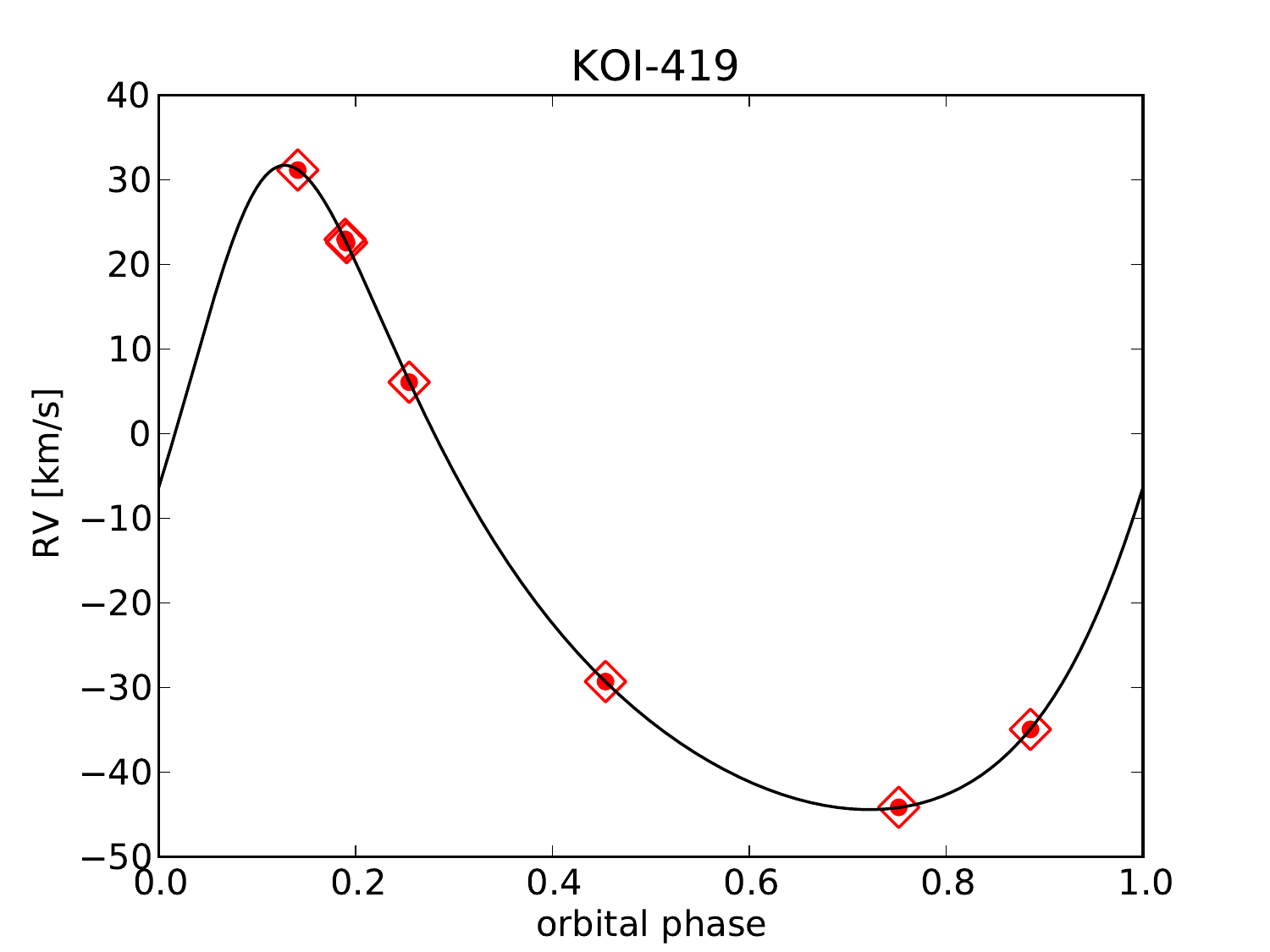}\\
\includegraphics[width=0.5\columnwidth]{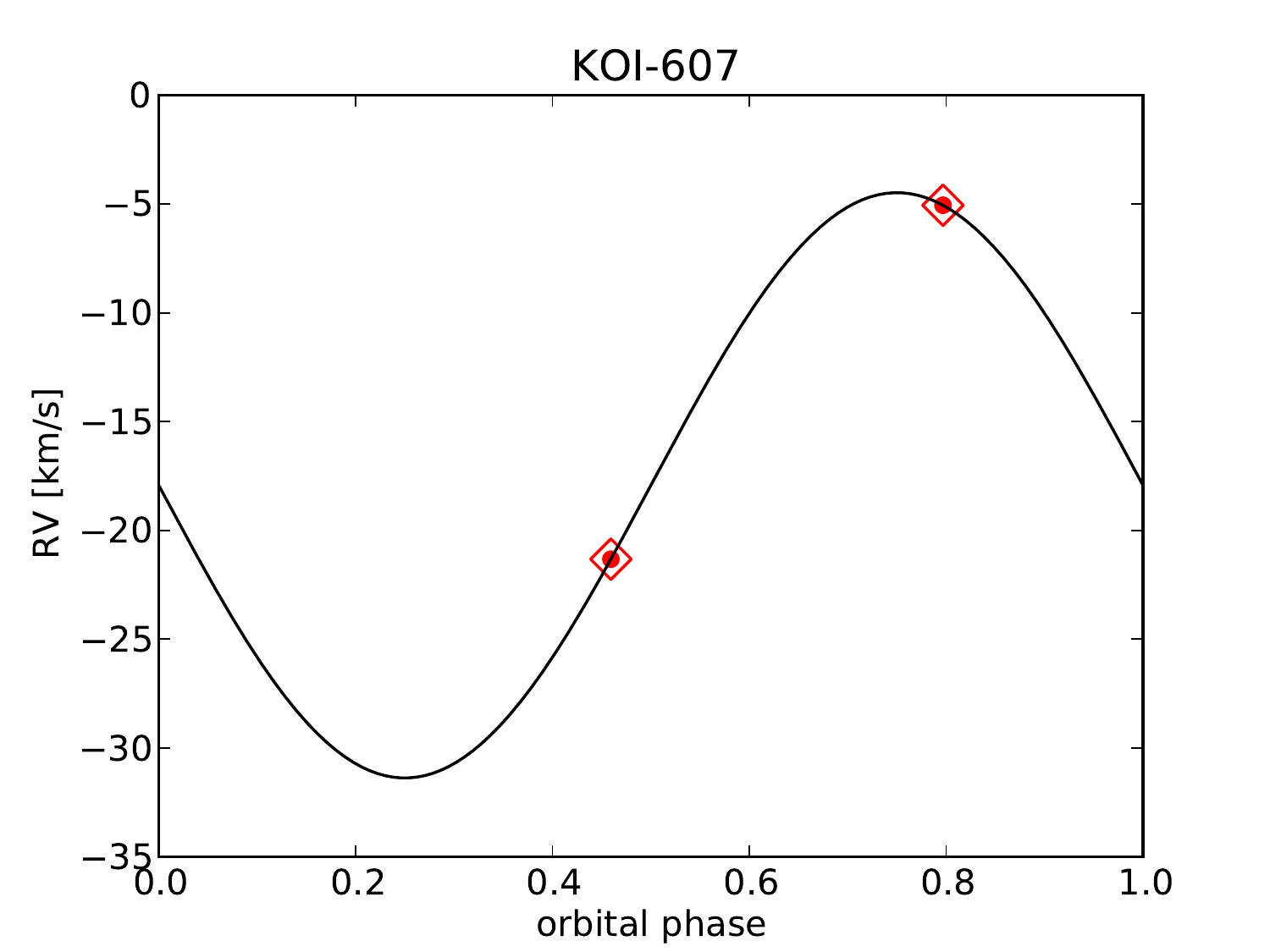} & \includegraphics[width=0.5\columnwidth]{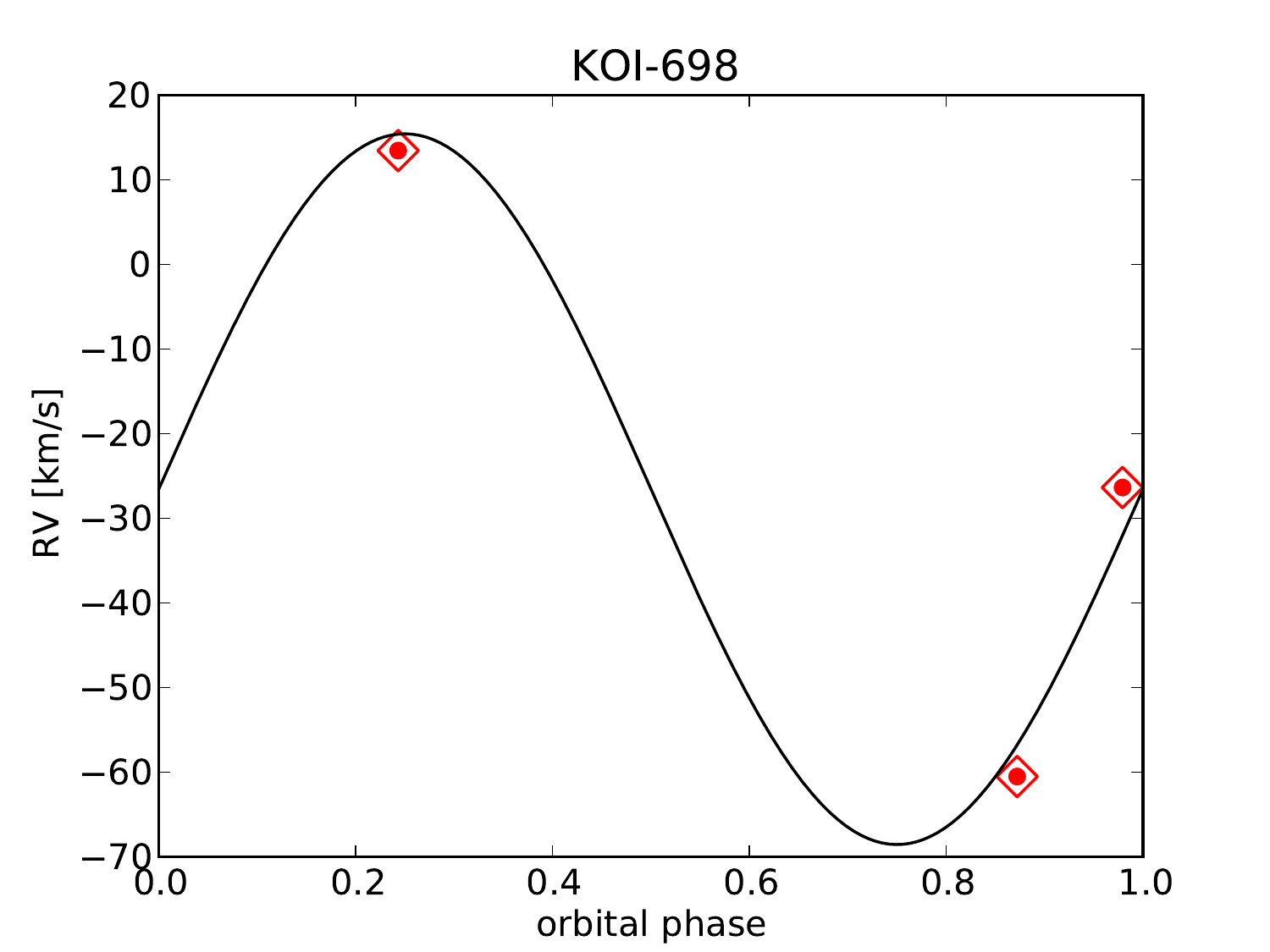}\\
\end{tabular}
\includegraphics[width=0.5\columnwidth]{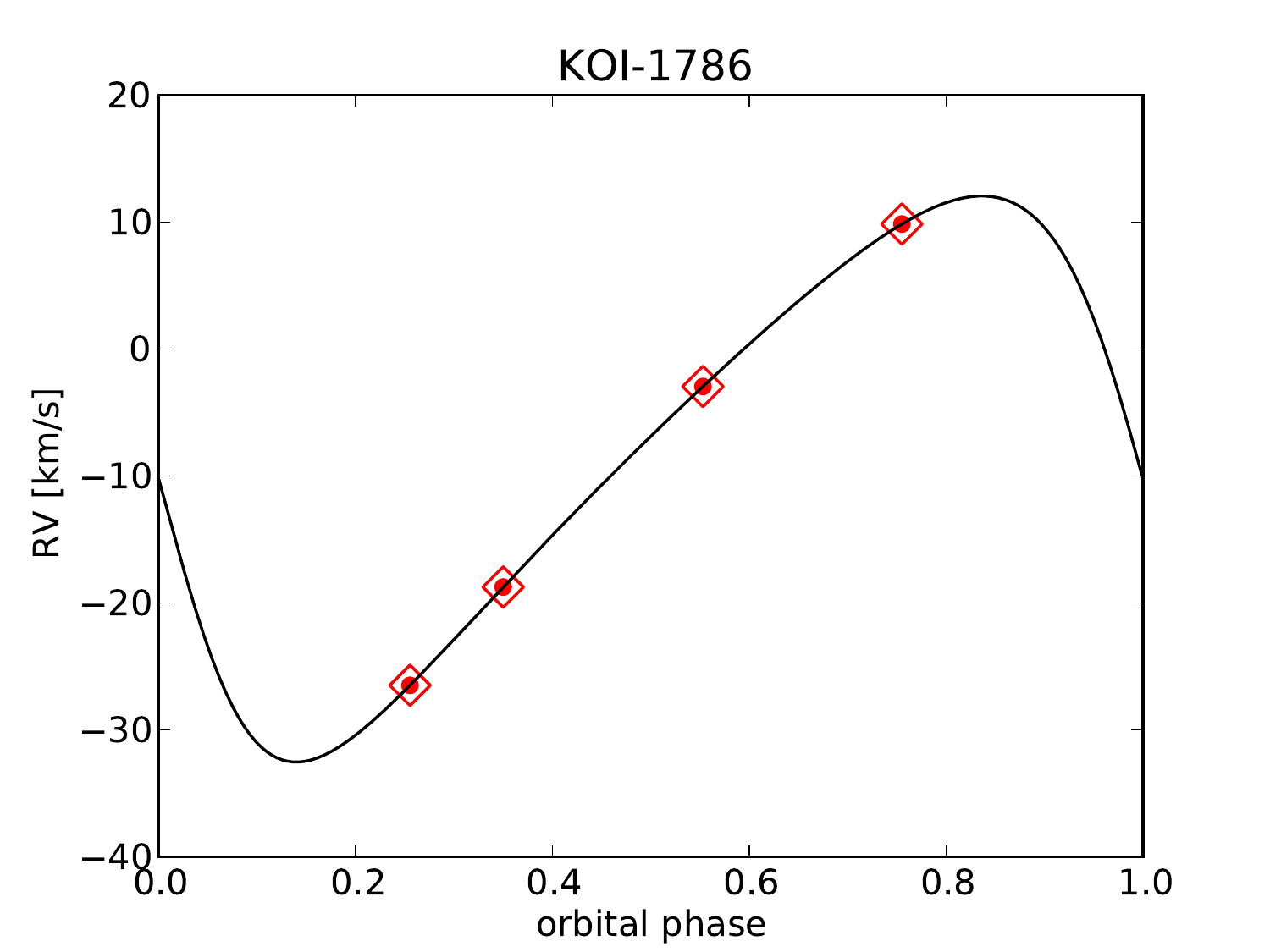}
\caption{Phase-folded radial velocity SOPHIE measurements of undiluted binaries where phase zero corresponds to the transit epoch. The black line displays the best circular or eccentric model.}
\label{figSB}
\end{center}
\vspace{-0.5cm}
\end{figure}

We observed the star KOI-1786 four times with SOPHIE. It hosts a transiting candidate with a period of 24.7 days. We measured a strong and eccentric ($e \sim 0.32$) radial velocity variation in phase with \textit{Kepler} ephemeris (see Table \ref{koi1786} and Fig. \ref{figSB}) that is caused by an eclipsing binary. Assuming a host star of 0.49 \Msun \citepalias{2012arXiv1202.5852B}, we found a mass for the transiting companion of $0.232 \pm 0.014$ \Msun. As for KOI-340.01, this planetary candidate is also listed in the \textit{Kepler} eclipsing binary catalog.

\subsection{Diluted binaries}

We also found several eclipsing binaries in our sample for which the transit is diluted by another star in the system (hierarchical system) or in foreground/background (blend):

\subsubsection{KOI-51.01}
KOI-51.01 is a candidate on an $\sim$ 10.4-day orbit. The digitalized sky survey shows three close-in stars within a nebulosity at the coordinates of this candidate. We roughly estimated the magnitude of each component seen on the POSSII F-DSS2 image using aperture photometry. We found that the transit would have a depth between 14\% and 21\%, depending on which star is transited. We therefore conclude that this candidate is a diluted binary. We note that this planetary candidate is also classified as a detached eclipsing binary in the \textit{Kepler} eclipsing binary catalog.

\subsubsection{KOI-190.01}

KOI-190.01 is an $\sim 12.3$-day-period candidate. We took two SOPHIE spectra at orbital phase 0.23 and 0.78 (see Table \ref{koi190}). The spectra revealed a double-line binary (see fig. \ref{figCCF}). The main component of this double-line presents a significant radial velocity variation in phase with the \textit{Kepler} ephemeris and with a semi-amplitude of $K = 14.186 \pm 0.059 \kms$ assuming a circular orbit. If this primary star has a mass of 1\Msun, its companion would have a mass of about 0.17\Msun. The second and fainter component also exhibits a radial velocity variation not in phase with the \textit{Kepler} ephemeris that is compatible with a drift of $\sim 42 \mathrm{m.s^{-1}.d^{-1}}$. Because the fainter component is not varying in anti-phase with the primary star with a larger amplitude, this system is probably a triple system with a low-mass star eclipsing the main component of a long-period binary. This candidate is therefore a diluted eclipsing binary, likely in a hierarchical triple system.

\subsubsection{KOI-418.01}
 We observed KOI-418.01 twice with SOPHIE, a candidate on a 22.4-day-period orbit. The observed CCFs revealed the presence of a blending companion (see fig. \ref{figCCF}). This blend scenario is confirmed by the bisector (see Table \ref{koi418}), which is clearly correlated with the radial velocities \citep{2009IAUS..253..129B}. We conclude that this candidate is a diluted eclipsing binary.

\subsubsection{KOI-425.01}
KOI-425.01 is an $\sim$5.4-day-period candidate that was observed twice with SOPHIE. As for KOI-190.01, the CCFs show two peaks (see fig. \ref{figCCF}), the main one of which is varying in phase with the \textit{Kepler} ephemeris. Assuming a circular orbit, we found a semi-amplitude of $K = 14.53 \pm 0.79 \kms$. Assuming a 1\Msun stellar mass, the transiting companion would have a mass of $\sim$0.13\Msun. The second and fainter peak presents no significant radial velocity variation. We therefore conclude that KOI-425 is a diluted eclipsing binary, likely in a triple system.

\subsubsection{KOI-609.01}
\begin{figure}[h!]
\begin{center}
\setlength{\tabcolsep}{0.0mm}
\begin{tabular}{cc}
\includegraphics[width=0.5\columnwidth]{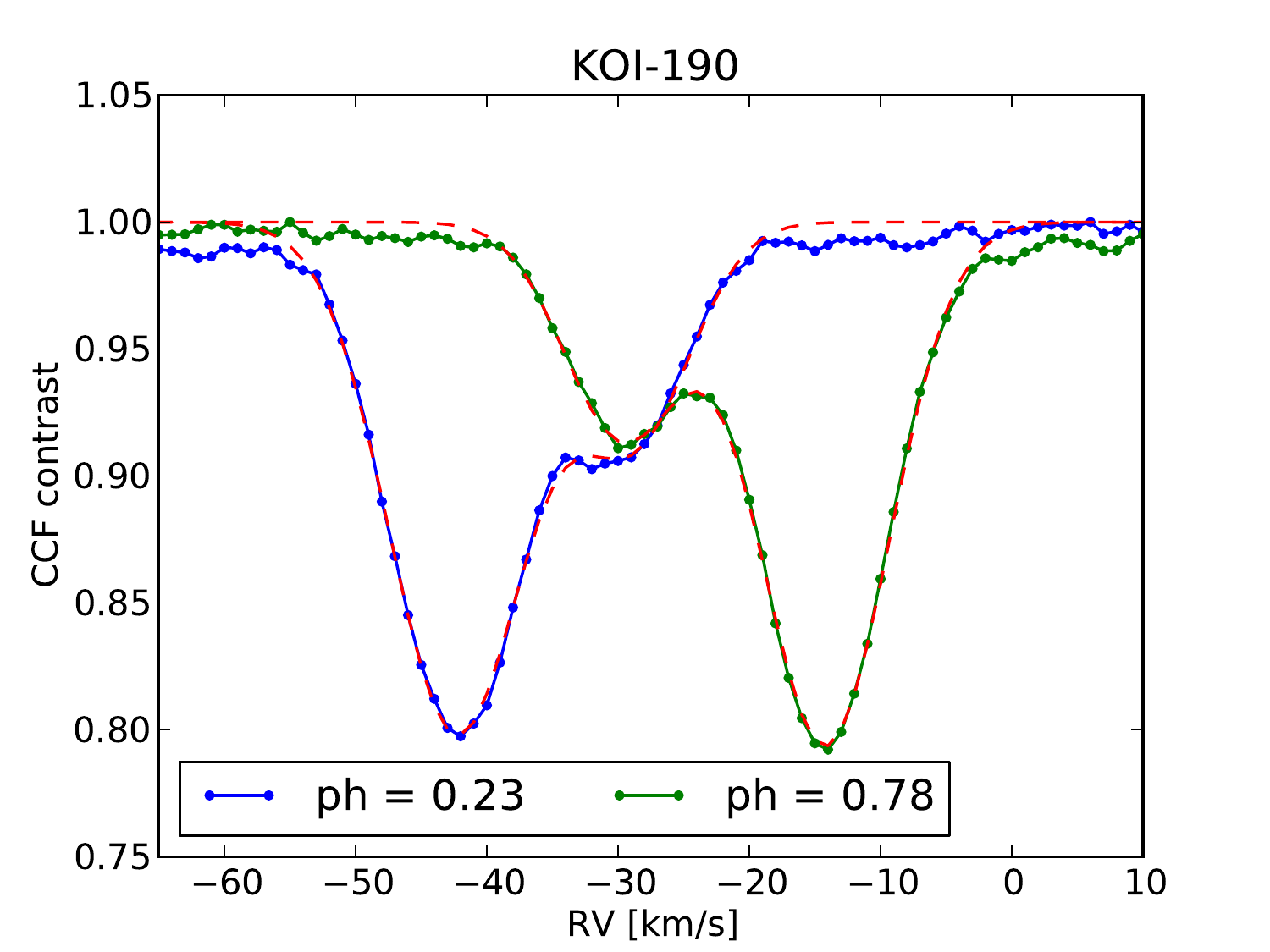} & \includegraphics[width=0.5\columnwidth]{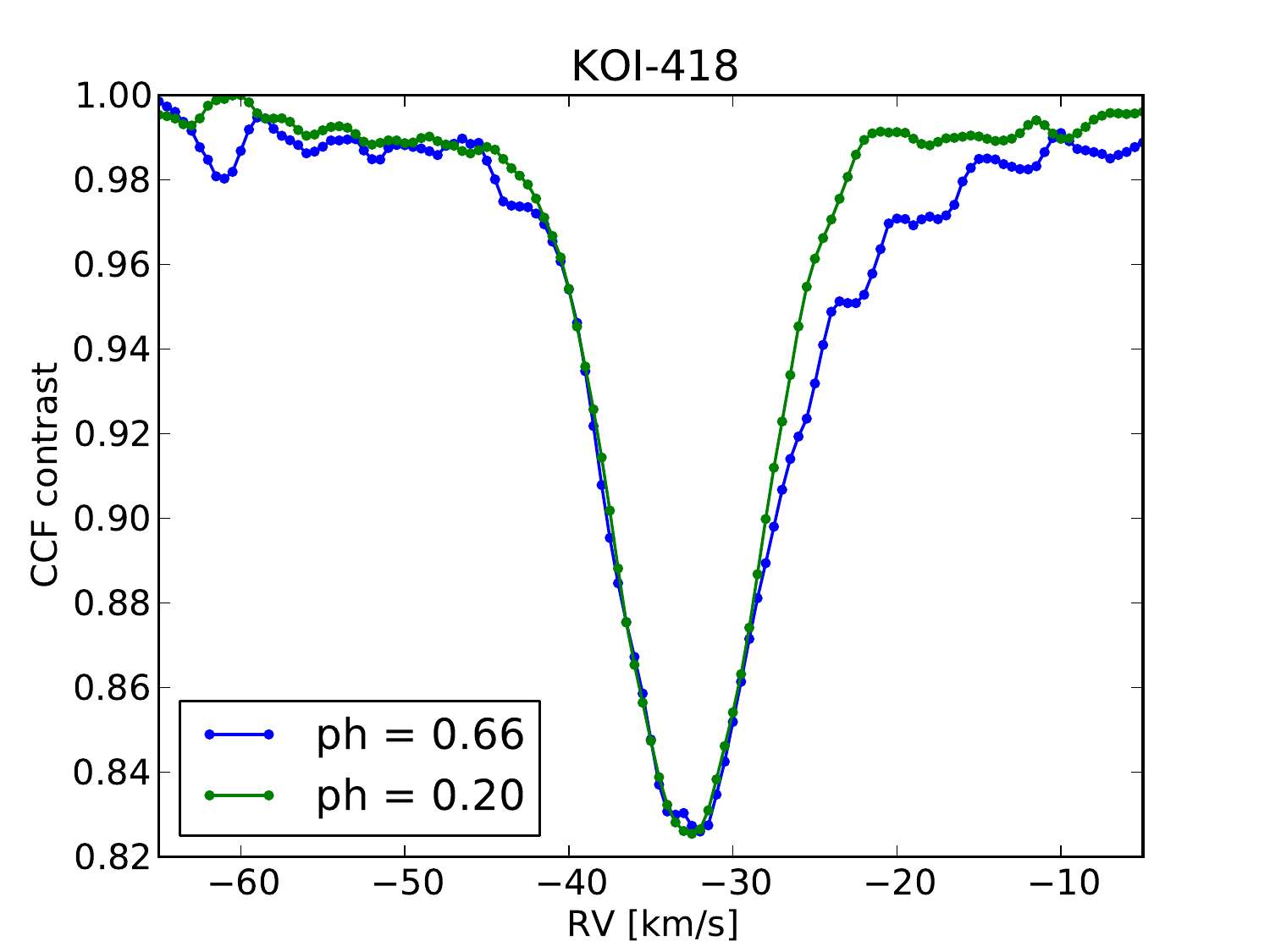}\\
\includegraphics[width=0.5\columnwidth]{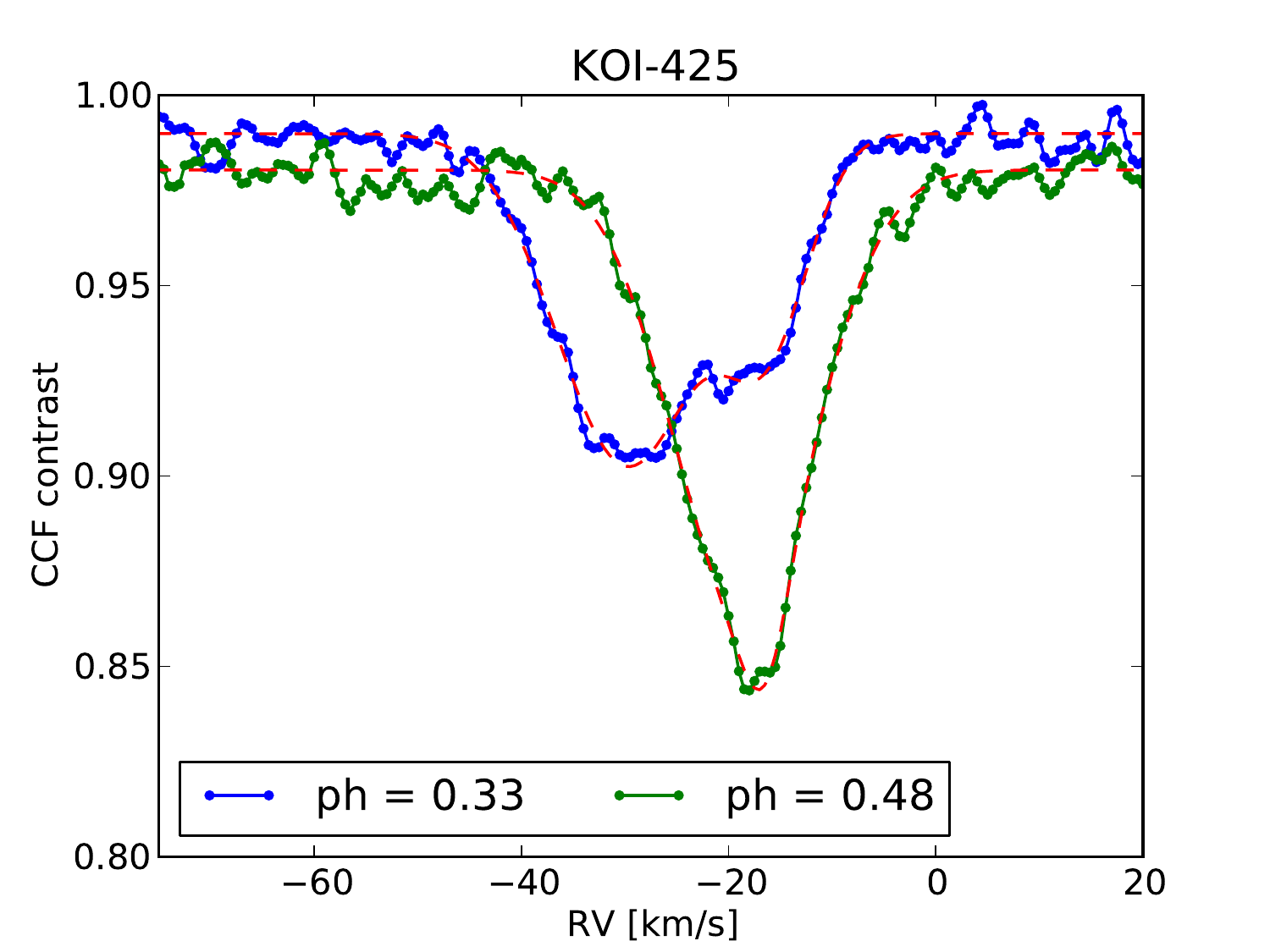} & \includegraphics[width=0.5\columnwidth]{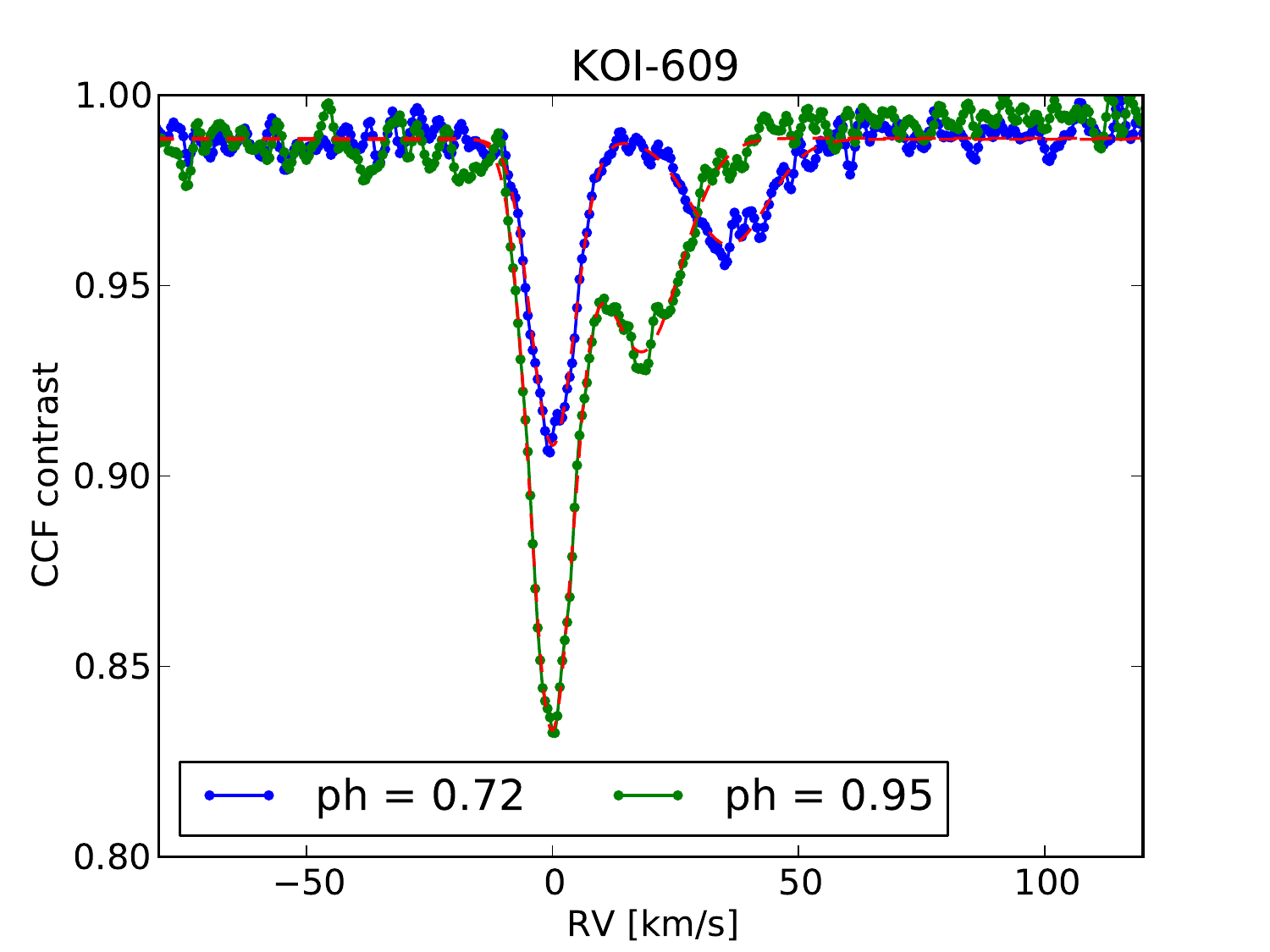}\\
\includegraphics[width=0.5\columnwidth]{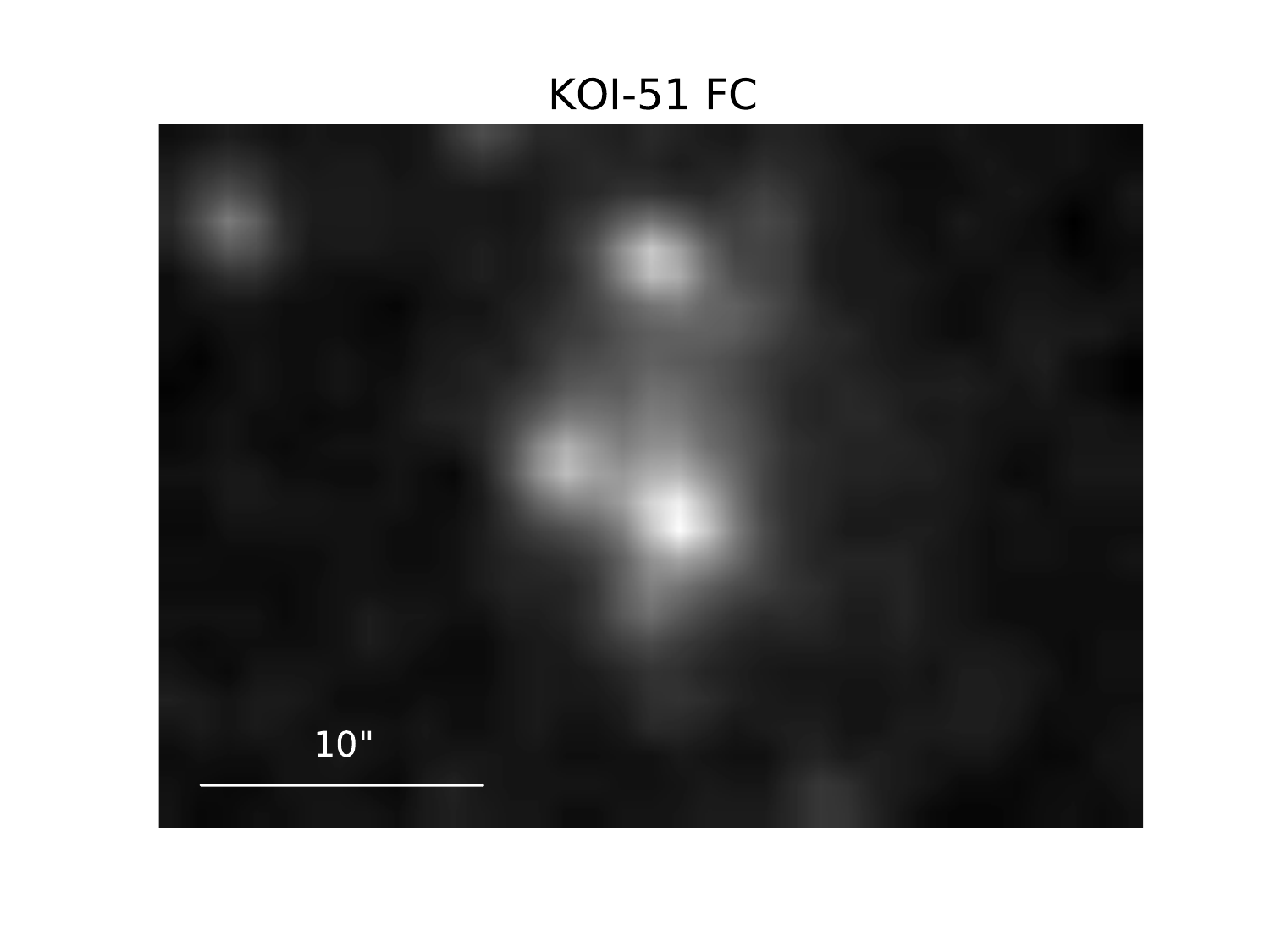} & \includegraphics[width=0.5\columnwidth]{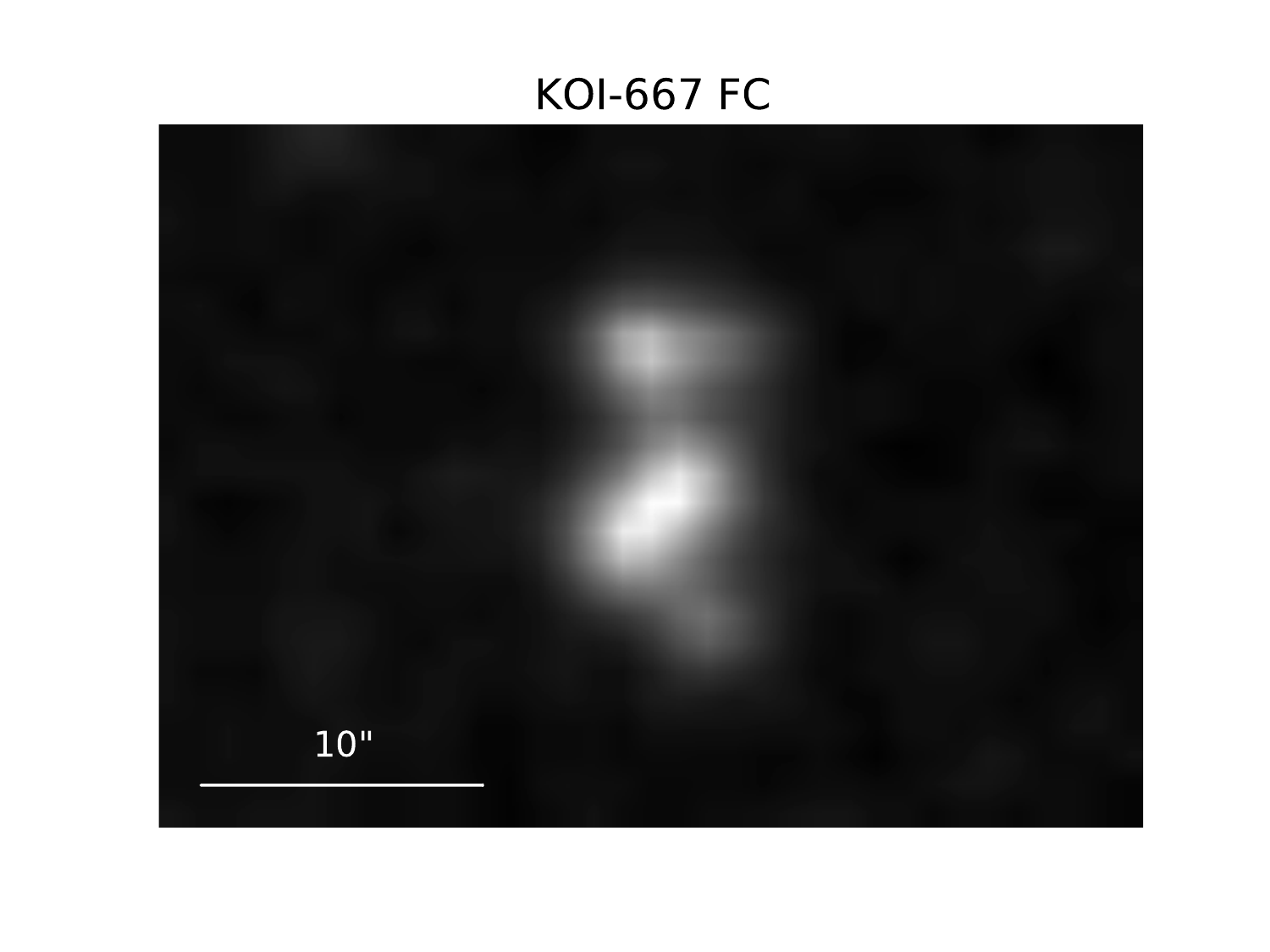}\\
\end{tabular}
\caption{CCFs of the diluted eclipsing binaries as a function of radial velocities. Blue and green dot-lines display the observed CCFs and the dashed red lines are the multi-Gaussian fit to the CCFs. The two lower panels are the SDSS finding charts (FC) of KOI-51.01 and KOI-667.01, which present several stars located within 20\arcsec x10\arcsec.}
\label{figCCF}
\end{center}
\vspace{-0.5cm}
\end{figure}

KOI-609.01 is a candidate that transits its host-star every $\sim$ 4.4 days. We observed it twice with SOPHIE. The observations revealed a double-line spectroscopic binary (see fig. \ref{figCCF}). We analyzed the radial velocity variation of both peaks in the CCF (see Table \ref{koi609}) and found that the primary peak presents no significant radial velocity variation. On the other hand, the secondary and fainter peak is varying in phase with the \textit{Kepler} ephemeris with a semi-amplitude of $K = 25.5 \pm 2.5 \kms$ if the orbit is circular. Assuming a host-star mass of 1\Msun, the companion of this secondary star would have a mass of $\sim$0.25 \Msun. Because the secondary peak is varying in phase with the \textit{Kepler} ephemeris, we conclude that KOI-609 could be either a triple system or an unresolved background eclipsing binary. We note that this candidate was also classified as a false positive by \citet{2011ApJS..197...12D}.

\subsubsection{KOI-667.01}
KOI-667.01 is a candidate on an $\sim$ 4.3-day orbit. As for KOI-51, the digitalized sky survey shows a diffuse object at the coordinates of this candidate with at least three blended stars. We also performed rough aperture photometry of the few stars seen in the POSSII F-DSS2 image. We found that the undiluted transit depth should be between 6\% and 12\%, depending on which star hosts the transit. We therefore concluded that this candidate is not a planet, but a diluted eclipsing binary.\\

\subsection{Unsolved cases}
\label{novar}
Some of the candidates present no significant radial velocity variations and are discussed below. Some of these candidates are members of the \textit{Kepler} eclipsing binary catalog \citep{2011AJ....141...83P} with twice the orbital period, which hints at a diluted equal-mass eclipsing binary. This it is no firm evidence of false positives because the planets KOI-135b \citep{bonomo2012} and KOI-206b (H\'ebrard et al., in prep.) are also in this list. We therefore considere them to be unsolved, pending new follow-up observations.

\subsubsection{KOI-12.01}
KOI-12.01 is an $\sim$17.9-day-period candidate orbiting a hot and fast rotating star with \teff$\sim 6400$K. We took two SOPHIE measurements at phases 0.31 and 0.74 (see Table \ref{koi12}) and did not detect any significant radial velocity variation at the level of $\sim640\ms$ assuming a circular orbit (see Fig. \ref{figNoVar}). With a period of about 18 days, the orbit may be eccentric. If this is the case, we may have missed the extremal phases. By fitting a rotational profile to the observed CCF, we found $\vsini  = 66 \pm 2 \kms$. Assuming a host star of 1.17\Msun \citepalias{2012arXiv1202.5852B}, we thus can put a 3-$\sigma$ upper limit on the mass of the companion of 26.7\Mjup. With only two points with such a large uncertainty, we cannot constrain any blend scenario by analyzing the correlation between bisector and RVs. No mask effect is seen above 1-$\sigma$. With a radius of about 1.12\Rjup, the transiting companion is still compatible with a planetary or low-mass brown dwarf scenario or with a blend scenario. We note that \citet{2011ApJS..197...12D} have detected an occultation depth for which they cannot conclude on a planetary or stellar origin. We conservatively considere this candidate to be unsolved.\\

\subsubsection{KOI-131.01}
We observed KOI-131.01 twice, an $\sim$ 5.0-day orbital-period candidate. The CCF revealed a fast rotating star with $\vsini = 27 \pm 1 \kms$. We found no significant radial velocity variation at the level of $\sim$ 800\ms  (see Fig. \ref{figNoVar}). We can therefore assume an upper limit on the mass of the expected companion that is lower than 14.3\Mjup assuming a circular orbit and a stellar mass of 1.28\Msun \citepalias{2012arXiv1202.5852B}. Our two measurements are not sufficiently accurate to allow a blend analysis using the bisector. No mask effect is seen above 1-$\sigma$. The planetary and blend scenarios are still compatible with our data. We note that this candidate is also classified as an eclipsing binary \citep{2011AJ....141...83P} with twice the orbital period.

\subsubsection{KOI-192.01}
KOI-192.01 is an $\sim$10.3-day-period candidate. We took two SOPHIE measurements at orbital phases 0.22 and 0.78 (see Table \ref{koi192}). The resulting radial velocities do not present any significant variation at a level of 23\ms  (see Fig. \ref{figNoVar}) assuming a circular orbit. If the orbit is slightly eccentric, we may not have observed KOI-192 at the extremal phases. From the CCF, we computed a $\vsini = 11 \pm 1 \kms$. By analysis of the spectra, we found a host star with \teff$ = 5976 \pm 165$ K, \logg = $4.46\pm0.15$ and \met $= -0.05 \pm 0.14$ dex in close agreement with the parameters published by \citetalias{2012arXiv1202.5852B}. This corresponds to a star with $M_{\star} = 1.01^{_{+0.13}}_{^{-0.11}}$ \Msun and $R_{\star} = 1.03 ^{_{+0.12}}_{^{-0.10}}$ \Rsun with an age of $3.5^{_{+6.3}}_{^{-1.4}}$ Gyr.\\

We therefore put a 3-$\sigma$ upper limit on the mass of the companion of 0.59\Mjup. We cannot constrain any blend scenario within 1-$\sigma$ with our two bisector measurements. No significant mask effect is seen up to the 1-$\sigma$ level. With a mass of less than 0.59\Mjup and a radius of $0.9\pm0.1$\Rjup, the transiting companion is still compatible with a Saturn-like planet or with a blend. High-constrast and high-resolution imaging \citep[e.g.][]{2012arXiv1205.5535A} would help to discard any potential background eclipsing binary within the exclusion radius of the centroid test \citep{2010ApJ...713L.103B}.

\subsubsection{KOI-197.01}
KOI-197.01 is a candidate in an $\sim$17.3-day-period orbit. We obtained twelve SOPHIE measurements (see Table \ref{koi197}). The resulting radial velocities do not present any significant variation at a level of 12\ms (see Fig. \ref{figNoVar}). We computed a \vsini of $11\pm1 \kms$. Our spectral analysis revealed a host star with \teff = $4995 \pm 126$ K, \logg $= 4.62Ê\pmÊ0.24$ and \met $= -0.11 \pm 0.06$ dex. This corresponds to an old star with $M_{\star} = 0.77 \pm 0.09$ \Msun and $R_{\star} = 0.74 \pm 0.08$ \Rsun.\\

We therefore put a 3-$\sigma$ upper limit on the mass of the companion of 0.27\Mjup. No significant bisector variation is seen in the data, nor mask effect within 1-$\sigma$. With a mass of less than 0.27 \Mjup and a radius of $0.65\pm0.07$ \Rjup, the transiting companion is still compatible with a Saturn-like planet or a blend. High-constrast and high-resolution imaging \citep[e.g.][]{2012arXiv1205.5535A} would also help to discard any potential background eclipsing binary within the exclusion radius of the centroid test \citep{2010ApJ...713L.103B}. We note that this candidate was classified as a false positive by \citet{2011ApJS..197...12D} based on the detection of an occultation depth significantly (4-$\sigma$ or more) deeper than expected. Because we rejected the scenario of an undiluted binary for this candidate but were not able to confirm the blend scenario, we conservatively consider this candidate to be unsolved.

\subsubsection{KOI-201.01}
KOI-201.01 is a candidate that transits its host star every $\sim$4.2 days. We observed this candidate twice with SOPHIE (see Table \ref{koi201}). The resulting radial velocities do not present any significant variation at a level of 33\ms. We found a \vsini of $9\pm1 \kms$. We found a host star with \teff $= 5526 \pm 231$ K, \logg $= 4.56 \pmÊ0.33$ and \met $= 0.28 \pm 0.15$ dex. This corresponds to a star with $M_{\star} = 1.09 ^{_{+0.13}}_{^{-0.16}}$ \Msun and $R_{\star} = 1.05 \pm 0.12$ \Rsun.\\

We therefore put a 3-$\sigma$ upper limit on the mass of the companion of 0.6\Mjup (see Fig. \ref{figNoVar}). From our two bisector measurements, we cannot constrain any blend. We found no mask effect above 1-$\sigma$ level. With a radius of 0.8 \Rjup and a mass of less than 0.6 \Mjup, neither the transiting planet scenario nor the blend scenario is discarded by our data. We note that this candidate is also in the \textit{Kepler} eclipsing binary catalog \citep{2011AJ....141...83P} with twice the orbital period.

\subsubsection{KOI-410.01}

As presented in \citet{bouchy2011}, we observed the candidate KOI-410.01 in 2010 twice with SOPHIE and found no significant variation that excludes at 3-$\sigma$ a planetary mass greater than 3.4\Mjup. \citet{2012arXiv1202.5852B} improved the parameters of this candidate compared to \citet{2011ApJ...736...19B} and found a radius ratio of $0.36151 \pm 1.36849$ by fitting a grazing eclipse to these 'V'-shaped events. Since we excluded any undiluted eclipsing binaries that could mimic a radius ratio this high, we suspect that this candidate is a blend without firm evidence.

\subsubsection{KOI-611.01}
\begin{figure*}[]
\begin{center}
\setlength{\tabcolsep}{5.0mm}
\begin{tabular}{c|c}
\includegraphics[width=0.85\columnwidth]{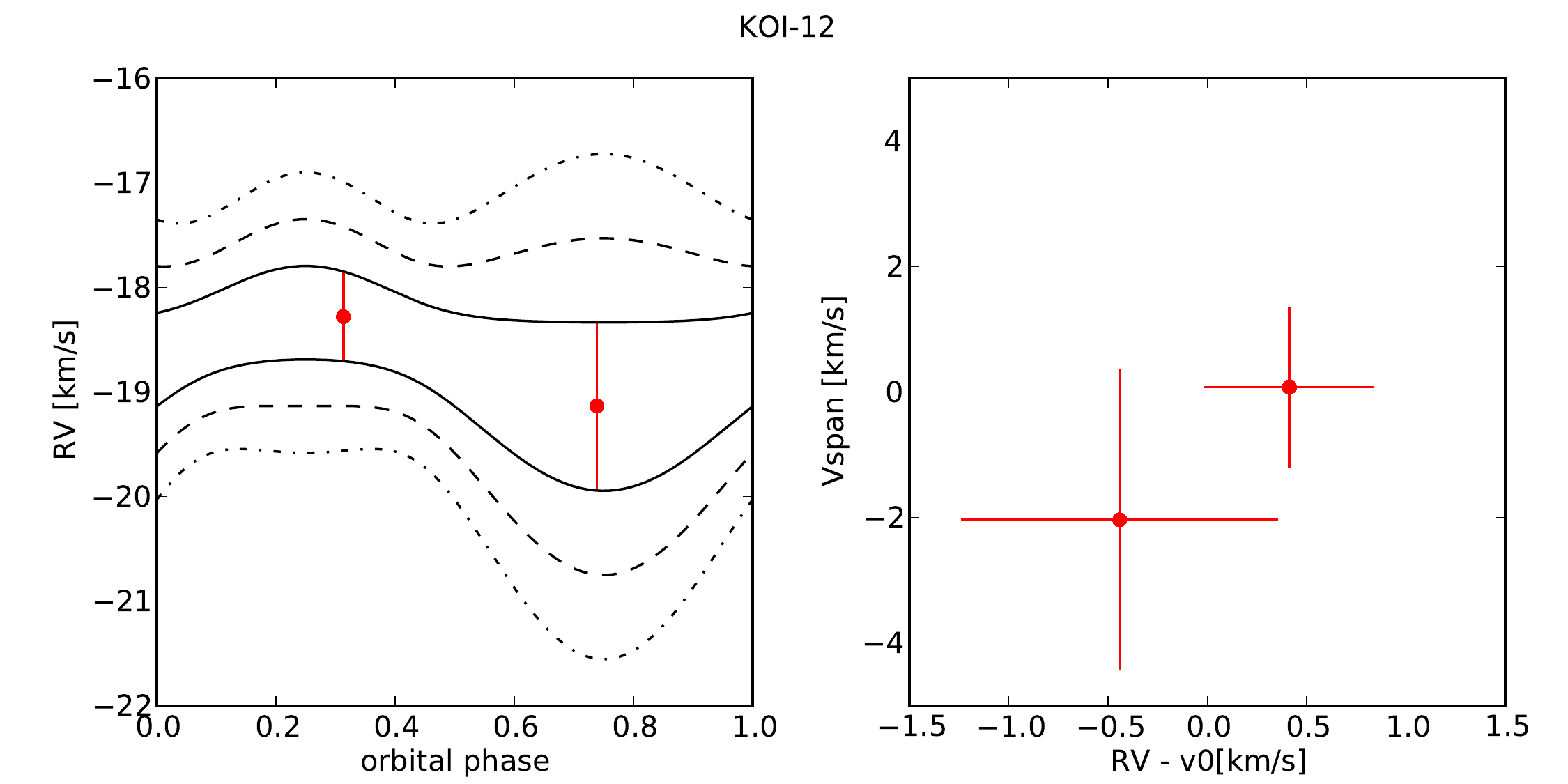} & \includegraphics[width=0.85\columnwidth]{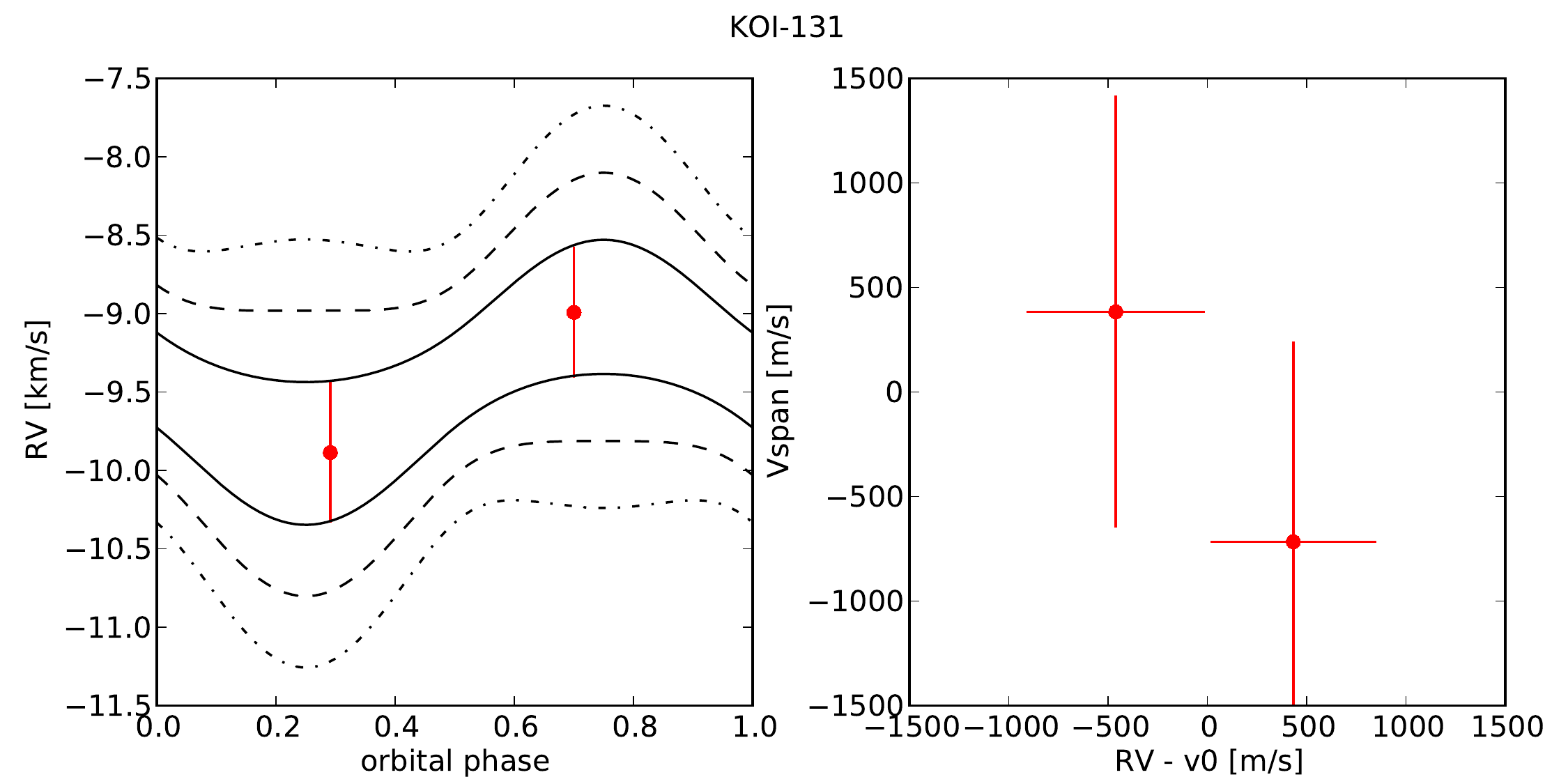}\\
\includegraphics[width=0.85\columnwidth]{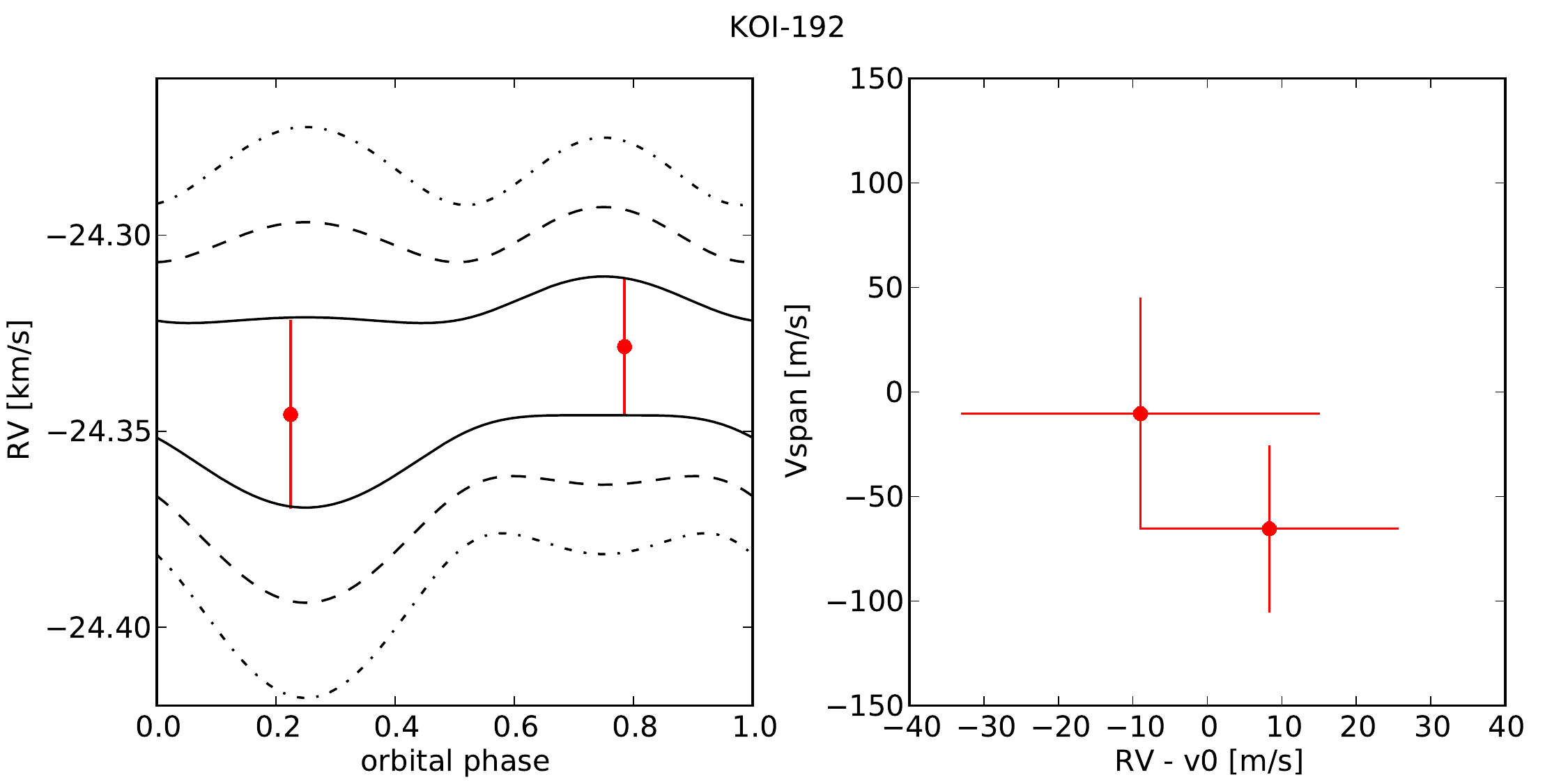} & \includegraphics[width=0.85\columnwidth]{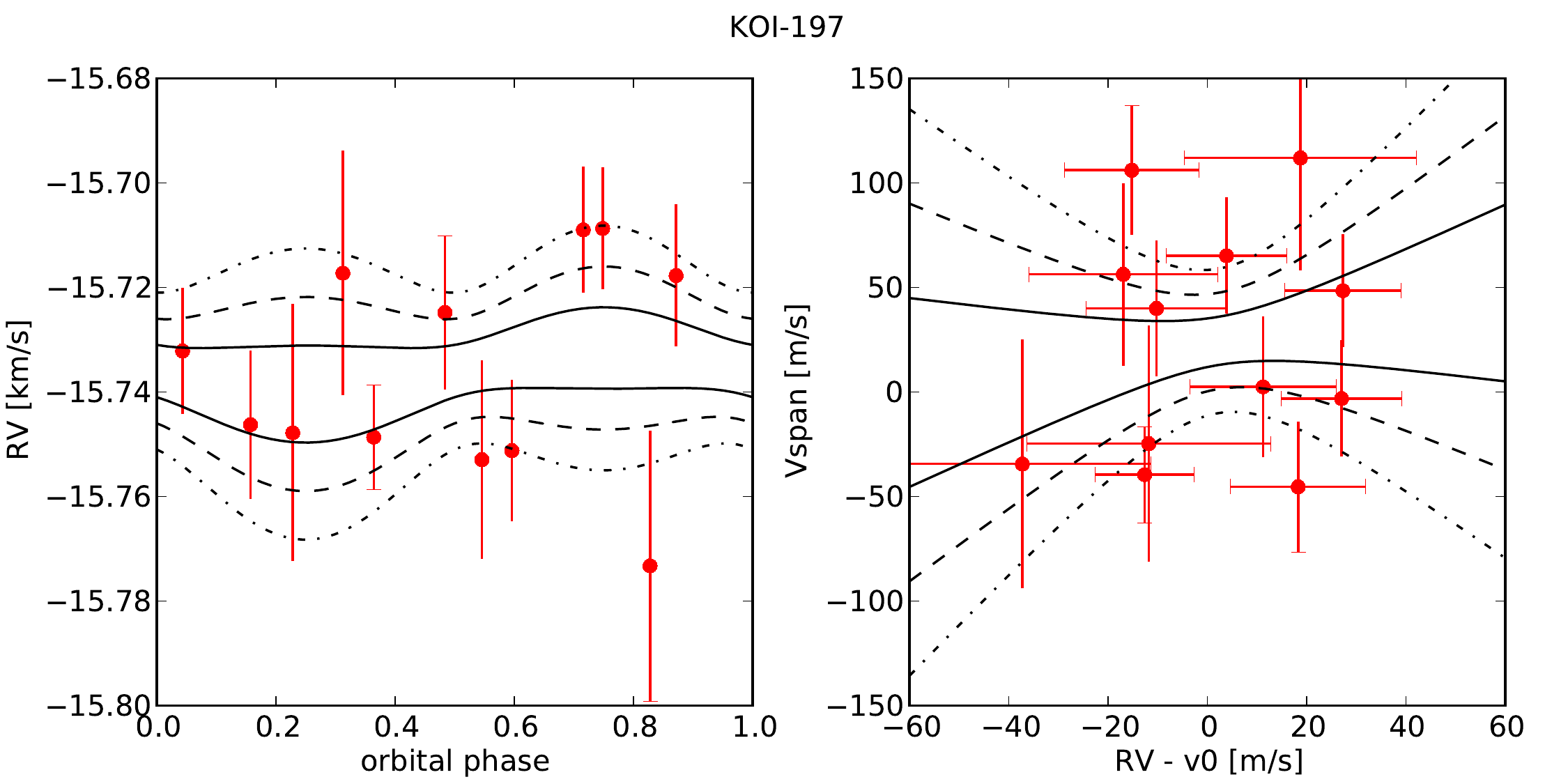}\\
\includegraphics[width=0.85\columnwidth]{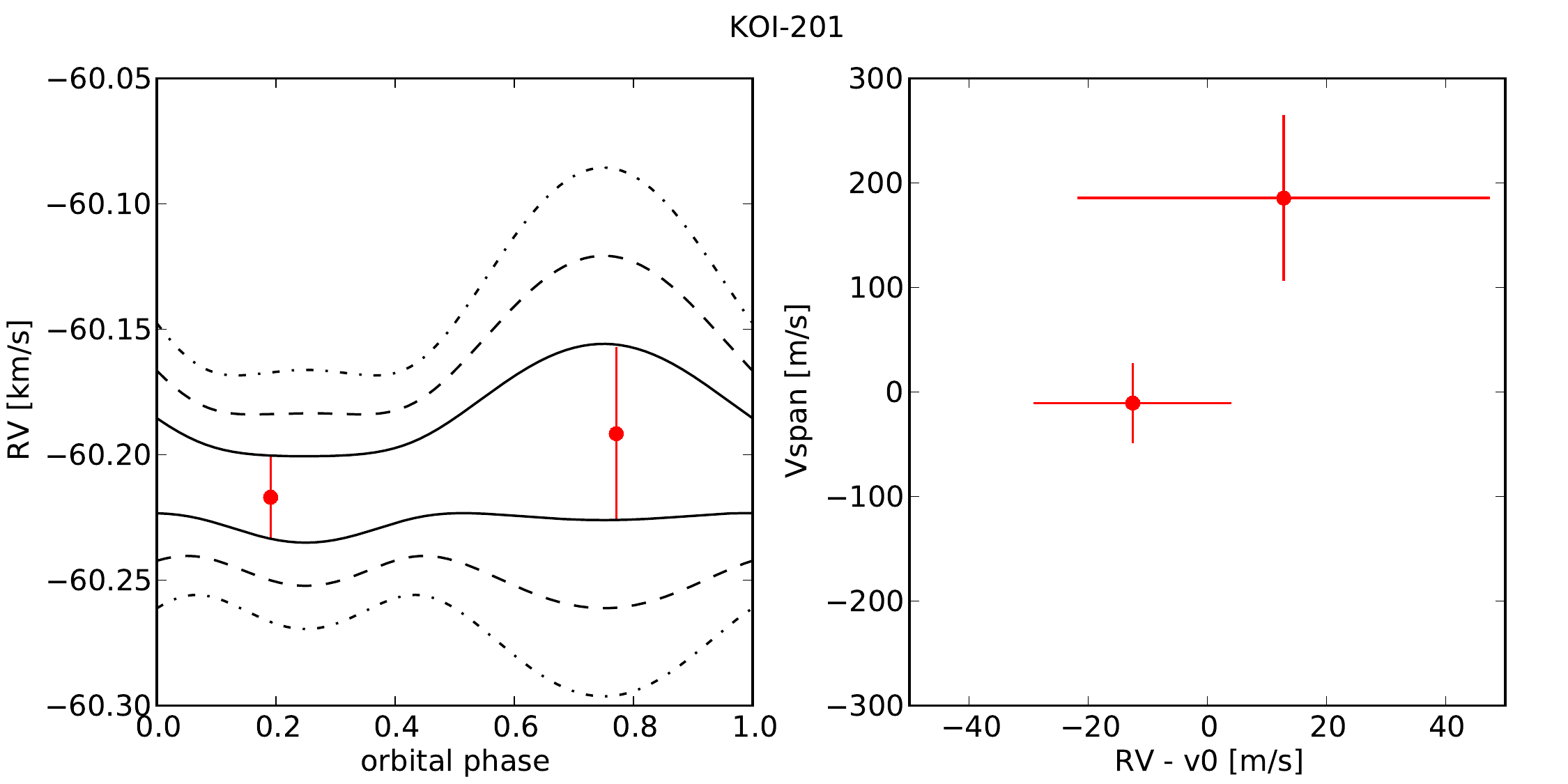} & \includegraphics[width=0.85\columnwidth]{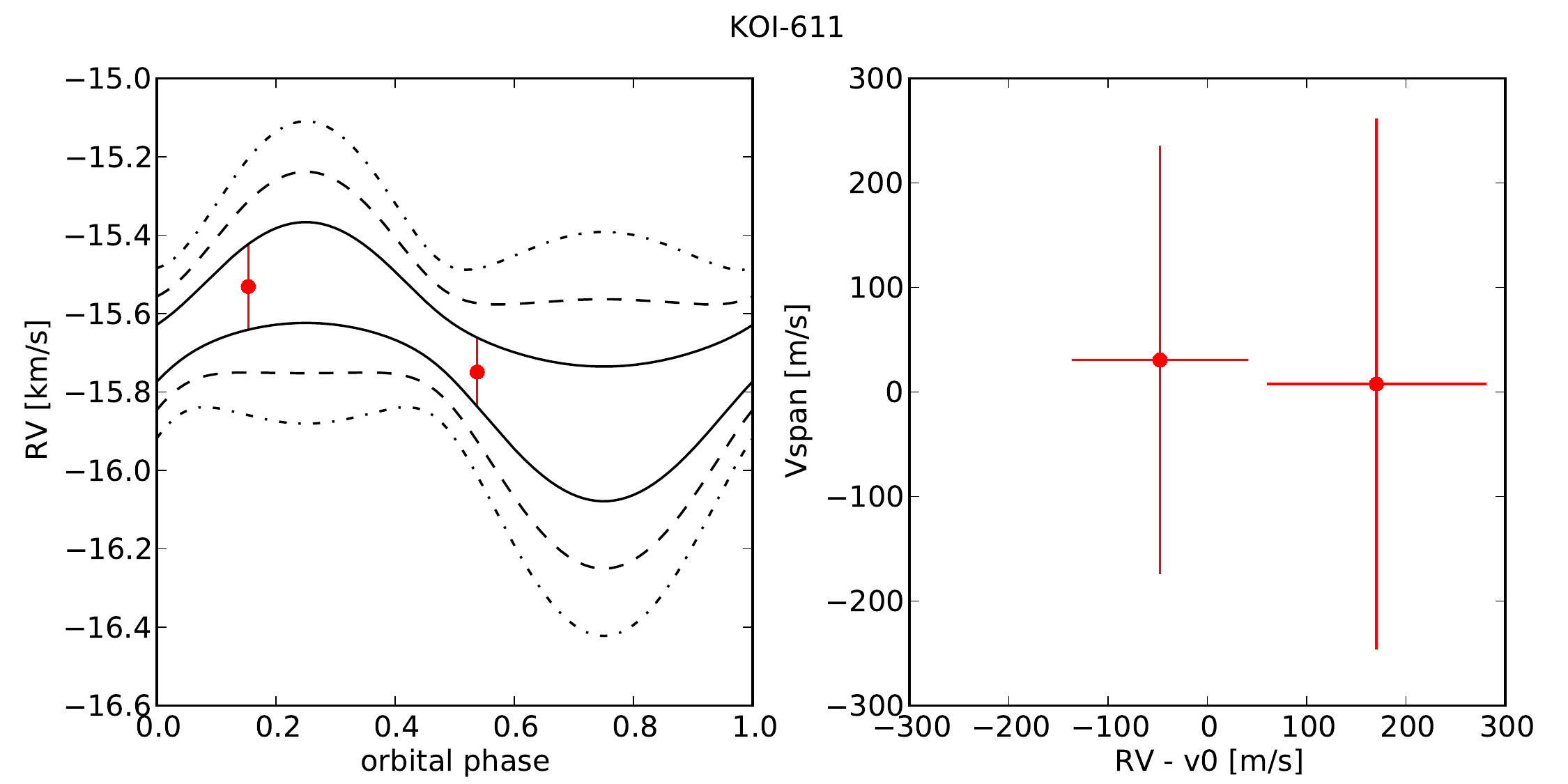}\\
\end{tabular}
\caption{Radial velocity \textit{(left panels)} and bisector \textit{(left panels)} SOPHIE measurements. Transits occur at phase zero. The straight line, dash-line and dash-dot lines represent the RV semi-amplitude and bisector variation limits at 1-$\sigma$, 2-$\sigma$ and 3-$\sigma$, respectively, assuming a circular orbit and a linear correlation between RV and bisector.}
\label{figNoVar}
\end{center}
\vspace{-0.5cm}
\end{figure*}

KOI-611.01 is a candidate that orbits its host star in $\sim$ 3.3 days. We took two SOPHIE spectra of this candidate that do not show any significant radial velocity variation at the level of $\sim$ 100\ms\, (see Table \ref{koi611} and Fig. \ref{figNoVar}). We found a \vsini of $17\pm1 \kms$. Assuming a host star mass of 1.09 \Msun, we can exclude all companions with a mass greater than 1.5 \Mjup with a 3-$\sigma$ confidence. Our two bisector measurements do not constrain any blend scenario. No mask effect is found in the data within 1-$\sigma$. With an expected radius of 0.65 \Rjup, the planetary scenario is still compatible with our data.

\subsubsection{Candidates not observed with SOPHIE}

We did not considered the remaining six candidates (KOI-22.01, KOI-63.01\footnote{KOI-63 was presented by Sanchis-Ojeda et al. at the First \textit{Kepler} Science Conference as a very likely misaligned planet.}, KOI-94.01\footnote{KOI-94.01 is a member of a multiple (4) system candidate and thus is very likely a planet.}, KOI-127.01, KOI-183.01 and KOI-214.01) for follow-up observations on SOPHIE since they were followed-up on other radial velocity facilities by the \textit{Kepler} follow-up team (Marcy, private comm.). We considered them as unsolved cases.

\section{The false-positive rate of \textit{Kepler} close-in giant candidates}
\label{section:FPP}


From the initial list of 46 KOIs selected, 20 planets have been discovered by various teams, seven are clearly undiluted eclipsing binaries or brown dwarfs, six are diluted eclipsing binaries, and the remaining 13 are still unsolved. This leads to a rate of 43.5\% $\pm$ 6.5\% of planets, 15.2\% $\pm$ 4.1\% of undiluted binaries, 13.0\% $\pm$ 4.3\% of diluted binaries, and finally 28.3\% $\pm$ 6.5\% of unsolved cases. The uncertainties were computed using 100,000 iterations of a bootstrap resampling technique. Each resampling consists of randomly selecting 46 candidates from the actual list of 46 objets observed, allowing for repetitions \citep[see][]{2010ApJS..190....1R}. The fractions of planets, false positives and unsolved cases are computed for each iteration. The resulting distributions are approximately normal for planets and unsolved cases, but not for the diluted and undiluted binaries, which are better fitted by a binomial distribution. The relatively low number of these false positives is not sufficient to reach the limit in which the binomial distribution resembles a Gaussian. In all cases, the quoted uncertainties correspond to the 68.3\% confidence region. We note that these uncertainties correspond to the statistical error only, and did not include any potential systematic source of error, such as potential misclassification of candidates.\\

\begin{figure}[h]
\begin{center}
\includegraphics[width=\columnwidth]{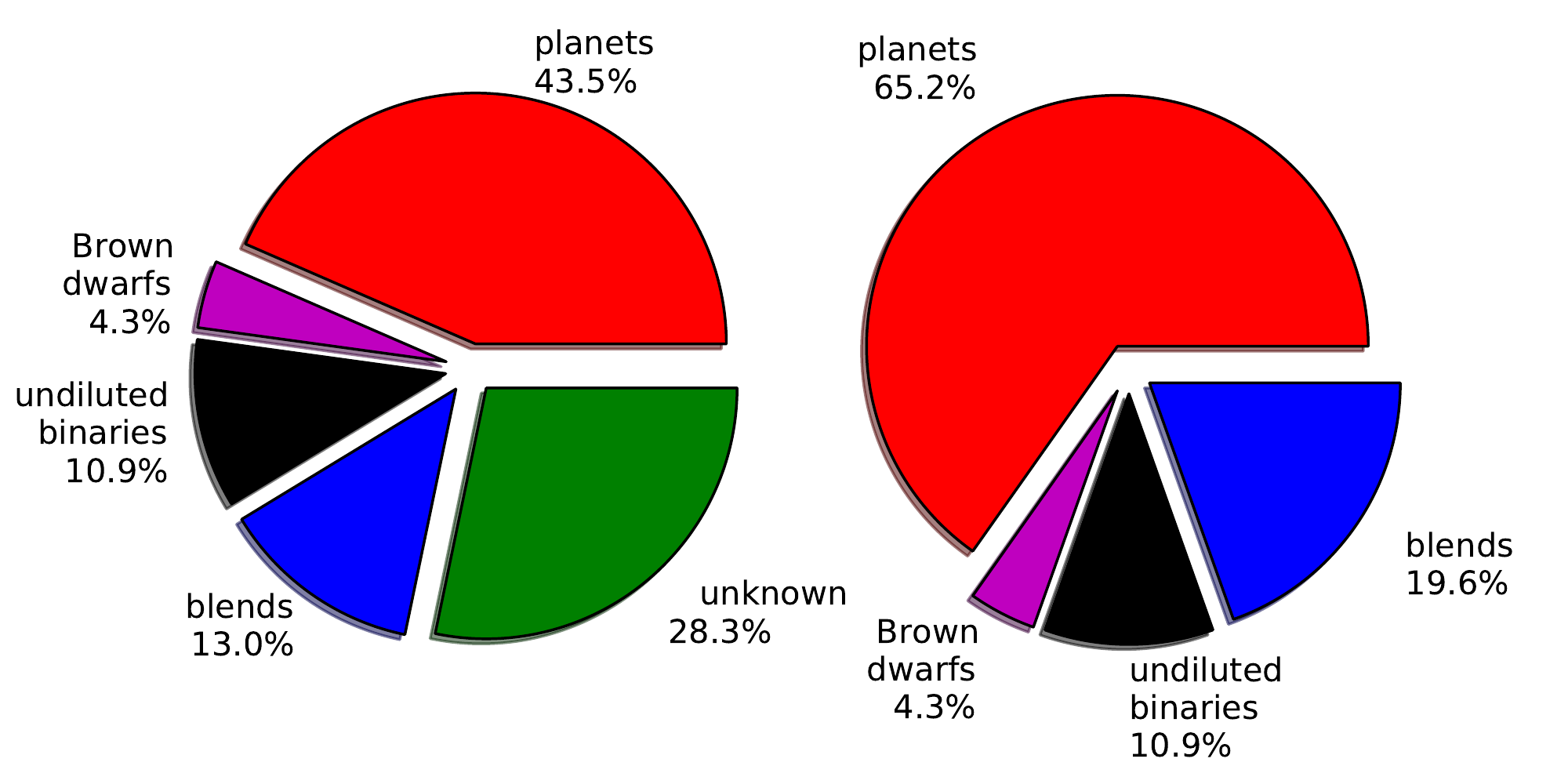}
\caption{Pie charts displaying the different ratio of each class of candidates. (\textit{left}) Raw pie chart from the results of observations. (\textit{right}) Pie chart assuming that the unsolved cases are either planets or blends with the same proportions as observed.}
\label{fig2}
\end{center}
\vspace{-0.5cm}
\end{figure}

The false-positive rate (FPR) of \textit{Kepler} giant candidates with orbital period shorter than 25 days and with a transit depth deeper than 0.4\% is thus between $28.3\%$ and $56.5\%$, depending on the true nature of the unsolved candidates (see Fig. \ref{fig2}, left pie chart). We can vouch that none of the unsolved case is an undiluted eclipsing binary because we would have detected a significant radial velocity variation. Consequently, we can assume that the true nature of the unsolved cases follows the same proportion of planets and blends as the observed one. This means that $76.9\%$ of them are low-mass planets and $23.1\%$ are diluted eclipsing binaries. We finally found that out of the 46 selected giant planet candidates 65.2\% $\pm$ 6.3\% are actual planets, 15.2\% $\pm$ 4.1\% are undiluted eclipsing binaries (including transiting brown dwarfs), and 19.6\% $\pm$ 6.5\% are blends. The \textit{Kepler} FPP for short-period giant planets is thus 34.8\% $\pm$ 6.3\%. We expected that if we had included the eight candidates with a vetting flag of 4, the FPP would be even higher. We note that if we focus on giant candidates with orbital periods shorter than 10 days, where the hot jupiter pile-up is expected, we find a lower FPR of $18.2\pm6.7$ \%.\\

\section{Discussions}

\subsection{The Morton \& Johnson (2011) estimation}
\label{MJ11}
The expected false-positive fraction for the Kepler candidates has been estimated by \citetalias{2011ApJ...738..170M} based on stellar population synthesis and galactic structure models. The authors also used the results of a comprehensive survey of stellar multiplicity of solar-type stars within 25 pc of the Sun \citep{2010ApJS..190....1R}. Their main result is that the expected false positive ratio of \textit{Kepler} candidates is below 10\%. More precisely, they found that for about 90\% of the candidates published by \citet{2011ApJ...736...19B}, the probability that they are false positives is below 10\%, and that about half have a FPP below 5\%. This result has motivated statistical analyses of planetary populations based on the \textit{Kepler} candidates alone \citep{2011arXiv1103.2541H}.\\

Our results seem to contradict the conclusions reached by \citetalias{2011ApJ...738..170M} for giant planet candidates. Indeed, our survey rejects a false-positive rate lower than 10\% with 99.99\% confidence level, considering the error on our FPR value. Moreover, when the FFPs of the individual candidates computed by \citetalias{2011ApJ...738..170M} are considered (see Table \ref{Saturn}), we found that the probability of having detected at least six false positives (our number of blended stellar systems, which are the only false positives considered in their analysis; see discussion below) is 0.4\%. This probability was computed using 100,000 Monte Carlo simulations based on our sample of candidates. In each iteration, each candidate is randomly decided to be a planet or a false positive, depending on its expected FPP (see Table \ref{Saturn}), and the resulting number of false positives ($n_\mathrm{FP}$) is recorded. The probability is obtained by integrating the distribution of $n_\mathrm{FP}$.\\

There are several reasons why the \citetalias{2011ApJ...738..170M} analysis may lead to an underestimation of the \textit{Kepler} FPP, at least for our sample of short-period giant planet candidates that represents about 2\% of all the \textit{Kepler} candidates. Some of these are evoked by the authors in the final section of \citetalias{2011ApJ...738..170M}.\\

Chief among them is the fact that undiluted binaries are not considered as a source of false positives. \citet{2011ApJ...738..170M} argue that this type of false positive can be effectively weeded out by a detailed analysis of the \textit{Kepler} photometry alone. However, we have found that more than 10\% of the followed-up candidates are actually low-mass-ratio binary stars, even excluding the two brown dwarfs reported here. This source of false positives is expected to be less important for smaller-radii candidates. However, as is clearly shown by the cases of KOI-419 and KOI-698, stellar companions in eccentric orbits and with relatively long periods can produce single-eclipse light curves, even for greater mass ratios. It is difficult to imagine how these candidates can be rejected from photometry alone if grazing transits are to be kept. After the submission of this paper \citet{2012arXiv1206.1568M} performed a new estimate of the FPP considering the undiluted binaries. His new analysis found a much higher FPP for most of our diluted and undiluted binaries.\\
 
However, even if we did not consider undiluted binaries, our survey has yielded at least six clear blended stellar systems, at most up to thirteen\footnote{This is not considering the six unsolved cases for which we have not performed follow-up, and of which two are most likely planets.}. Our best estimate for the fraction of blended stellar systems is 19.6\% $\pm$ 6.5\%. We are therefore led to conclude that \citetalias{2011ApJ...738..170M} underestimated this sample of candidates. The first possible reason that comes to mind is an underestimation of the stellar density in the direction of the \textit{Kepler} field. The authors used the TRILEGAL code of stellar population and Galactic structure \citep{2005A&A...436..895G}. The star count in TRILEGAL has been reported to show discrepancies smaller than 30 per-cent with a variety of stellar surveys for most of the sky, but it exhibits ``major discrepancies'' with fields at galactic latitude $\lesssim$ 10 degrees \citep{2005A&A...436..895G}. About 24\% of the \textit{Kepler} planetary candidate hosts would be affected by this effect. Although some of the simulated fields used to test TRILEGAL show a \emph{larger} number of stars than observations \citep{2005A&A...436..895G}, we believe that a detailed analysis of the star count yield is warranted. This task is beyond the scope of this paper.\\

A possible underestimated source of false positives in the \citetalias{2011ApJ...738..170M} analysis are blended equal-mass eclipsing binaries, for which the difference in depth of the diluted primary and secondary eclipses will be too small to be detected by \textit{Kepler} photometry. Since the mass ratio distribution of binary systems has a peak at $q \sim 1$ and short-period (P $<$ 100 days; i.e. those with the highest probability of eclipsing) binary systems tend to have higher mass ratios \citep{2010ApJS..190....1R}, this type of eclipsing binaries might contribute significantly to the number of false positives. We note that none of our false positives appears to be a diluted equal-mass binary. However, we note that this source of false positives is expected to be more common for smaller planet candidates since odd/even depth difference as well as transit shape are less significant.\\

Another possibility is an incorrect assumption for the planet radius distribution. \citet{2011ApJ...738..170M} consider a continuous power law  that increases toward small radii ($dN/dR_p \propto R_p^{-2}$), but warn that if this is not so, false positives might be twice as numerous.\\

Finally, a crucial factor in the number of blending systems affecting a given target is the area around it within which blends can reside. This area depends on the precision of the photocenter position, which \textit{Kepler} measures for all of its candidates. \citet{2011ApJ...738..170M} assumed a scaling law for the precision of the photo-center (their eq.\ 19) based on the host star magnitude and the depth of the observed transits. To be conservative, they assumed a minimum ``radius of 2\arcsec\, inside which a blend might reside''. If this blend exclusion radius is underestimated, it will definitely underestimate the proportion of potential background eclipsing binaries.\\

Even if this minimum exclusion radius of 2\arcsec, i.e. equivalent to one pixel of \textit{Kepler}, seems to be quite conservative, the number of transits observed for a given candidate should also be taken into account, since the position of the photo-center out of transit must be compared with the position in transit. Therefore, short-period candidates exhibiting many of transits should produce more precise measurements of the centroid shift. Since the scaling law is based on the measurements of a single very short-period planet (Kepler-10 b, P = 0.84 days), we believe that \citetalias{2011ApJ...738..170M} might overestimate the ability of \textit{Kepler} to identify blended stellar systems from astrometric measurements. \\

As mentioned by \citetalias{2011ApJ...738..170M}, they assumed candidates to have passed all vetting procedures and that there is no clear V-shape transit or deep secondary transit, which is obviously the case \citep[e.g.][]{bouchy2011, 2011ApJS..197...12D, 2012AJ....143...39C}.

\subsection{Comparison with other FPP estimations}
\label{otherFPP}

Our value for the FPP agrees better with the one estimated by \citet{2011ApJ...736...19B}, who estimated a FPP of  $<20\%$ and $<40\%$ for KOIs with a vetting flag of 2 and 3, respectively. \citet{2010arXiv1001.0352G} discussed 21 good candidates with magnitude brighter
than 14 followed-up. They found five planets (24\%), eight rejected (38\%) and eight without conclusion. They then claimed a FPP in between 38\% and 76\%, which is roughly compatible. Based on the occultation depth found in the \textit{Kepler} light-curves, \citet{2011ApJS..197...12D} and \citet{2012AJ....143...39C} found a FPP of 14\% and 11\%, respectively. This method shows its limitations since \citet{2011ApJS..197...12D} did not reject the four candidates that we found to be clearly diluted binaries (KOI-190.01 and KOI-425.01) or undiluted eclipsing binaries (KOI-205.01 and KOI-698.01). However, they classified KOI-609 as a false positive which we confirmed to be a diluted eclipsing binary, and KOI-197, for which we were not able to constrain the planetary or blend scenarios.

\subsection{Extrapolation to longer orbital period giant-planet candidates}
\label{longPeriod}

\begin{figure}[t]
\begin{center}
\includegraphics[width=\columnwidth]{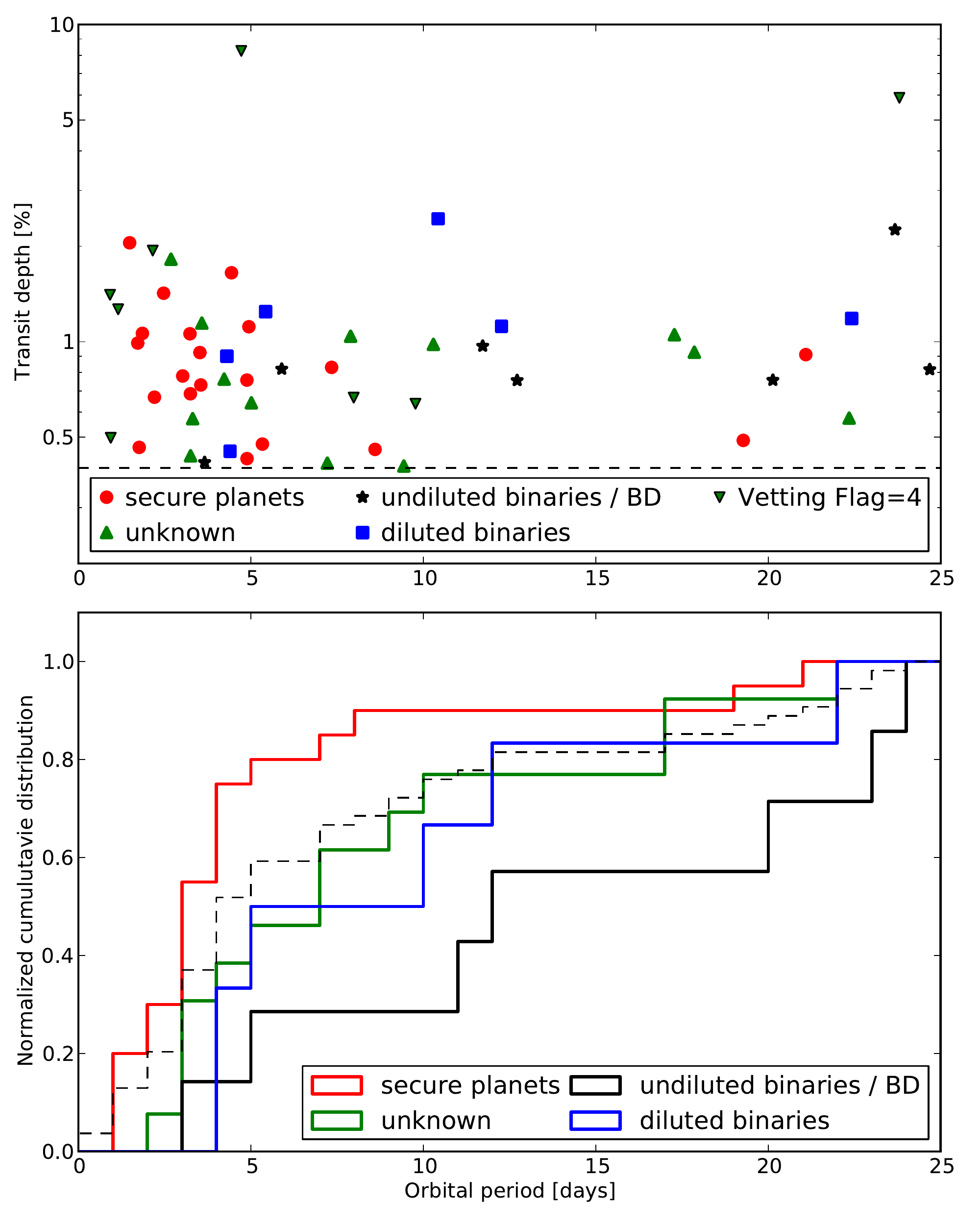}
\caption{(Top panel) Transit depth of the 46 selected KOIs as a function of their orbital period. The different marks represent the different nature of the candidate. The dashed line represents the cut in transit depth applied to this selection. (Lower panel) Normalized cumulative distributions of the orbital period of the different classes of candidates. The dashed black line displays the normalized cumulative distribution of the 46 selected candidates.}
\label{fig1}
\end{center}
\vspace{-0.5cm}
\end{figure}

\begin{figure}[h]
\begin{center}
\includegraphics[width=\columnwidth]{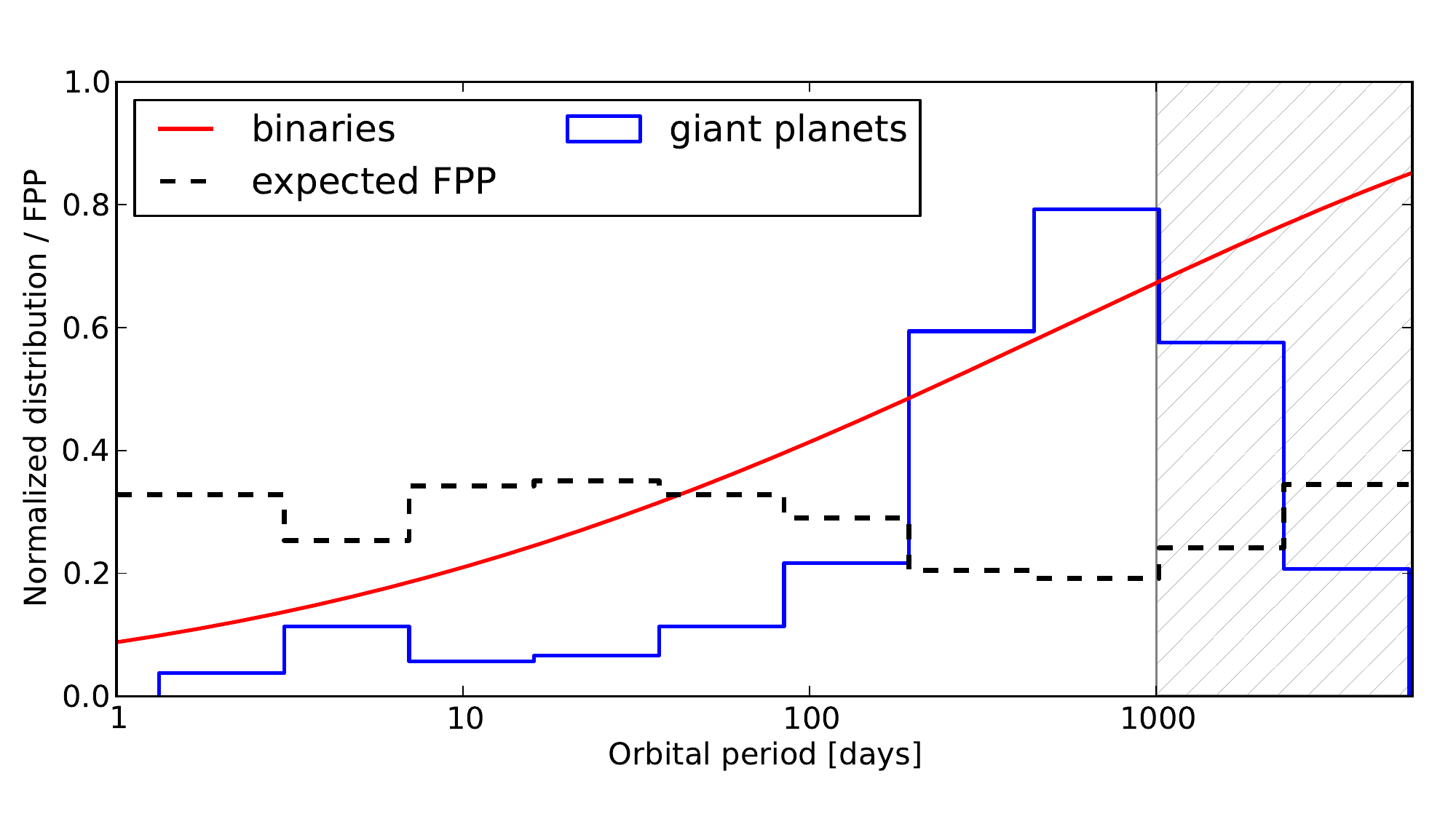}
\caption{Normalized distribution of 294 giant extrasolar planets discovered by the radial velocity technique to date with a mass greater than 0.5 \Mjup (blue line). The normalized distribution of binaries within a 3000-day-period from \citet{2010ApJS..190....1R} is displayed with the red line. The expected FPP distribution for giant \textit{Kepler} candidates is overplotted with the dashed black line. The hatched region represents planet orbital periods that we considered to be affected by observational bias. The FPP in this region is therefore expected to be overestimated.}
\label{rFPP}
\end{center}
\vspace{-0.5cm}
\end{figure}

Figure \ref{fig1} displays the cumulative period distributions of planets (red line) and false positives (blue and black lines) for our candidate selection. The cumulative distribution of all candidates is shown as a dashed line. These distributions confirm the existence of a pile-up of giant planets at very short orbital period ($\sim 3$ days). In contrast, distributions of both diluted and undiluted binaries are relatively flat over the observed period range. It is expected from the radial velocity surveys and from the binary population \citep{2010ApJS..190....1R} that both populations have a different period distribution (see also Fig. \ref{rFPP}). To test if our results agree with such a different distribution, we performed Kolmogorov -- Smirnov tests on the distribution of planets and false positives. We can reject the hypothesis that planets and undiluted binaries, or planets and blends, have the same period distribution with a probability of more than 97.5\%. We can also reject that planets and false positives have the same period distribution with a probability of 99\%. We thus found a different period distribution for planets and false positives. Indeed, no pile-up at very short period is expected for binaries, which is the case for giant planets.\\

These results are compatible with the relative distribution of planet orbital periods and binary orbital periods (see Fig. \ref{rFPP}). Indeed, we can express the FPP as follows:

\begin{eqnarray}
\label{FPPeq}
\mathrm{FPP}(P) = 1 - \frac{\pi_{pl} (P)}{\pi_{pl}(P)+\pi_{\star}(P)},
\end{eqnarray}
where $\pi_{pl}(P)$ and $\pi_{\star}(P)$ are the probabilities of having a planetary companion or a stellar companion (respectively) diluted or not for a given orbital period $P$. $\pi_{\star}(P)$ is the sum of probabilities to have both a diluted or an undiluted binary:

\begin{equation}
\label{probstar}
\pi_{\star}(P) = \pi_{BB}(P) + \pi_{PT}(P) + \pi_{SB}(P),
\end{equation}
where $\pi_{BB}(P)$, $\pi_{PT}(P)$ and $\pi_{SB}(P)$ are the respective probabilities of having a background binary, a physical triple system or a spectroscopic binary.
If we assume that all these binaries follow the same period distribution, we can reduce equation \ref{probstar} to $\pi_{\star}(P) \propto \pi_{SB}(P)$ and equation \ref{FPPeq} to 

\begin{equation}
\label{FPPeq2}
\mathrm{FPP}(P) \propto 1 - \frac{1}{1+\frac{\pi_{SB}(P)}{\pi_{pl}(P)}}.
\end{equation}

Figure \ref{rFPP} displays period distributions of giant planetary companions detected by radial velocity (blue line) and stellar companions (red line). We adopted the \citet{2010ApJS..190....1R} result for the binary period-distribution: $\log_{10}(P\ [\textrm{d}]) = \mathcal{N}\left(5.03, 2.28\right)$, where $\mathcal{N}(\mu, \sigma^{2})$ is the normal distribution centered on $\mu$ with a standard deviation of $\sigma$. We note that this binary period distribution was calibrated only for binaries and triple systems of solar-type primary stars in the solar neighborhood. We may expect that distribution of binaries is different when dealing with non solar-type stars. The estimated FPP for \textit{Kepler} giant planet candidates (fig. \ref{rFPP} dashed line) was computed using eq. \ref{FPPeq2} and calibrated to have FPP =  34.8\% when considering the same period range as our selected candidates period distribution.\\

We found that while the number of binaries increases with orbital period up to about 300 years, the number of detected giant planets decreases for periods between $\sim 10$ and $\sim 200$ days. This so-called ``period valley'' detected by radial velocity surveys cannot be explained with an observational bias since it is easier to detect planets with short orbital period than planets with orbital periods of a few years. This valley implies that the FPP should be constant over periods of less than 200 days. In contrast, candidates with orbital periods longer than $\sim 200$ days might have a FPP lower than 20\%. We consider that the distribution of planets with orbital periods longer than about three years is underestimated due to observational bias and that the respective FPP should be lower.\\

\subsection{Impact on exoplanet statistics}
\label{exoStat}

\citet{2011arXiv1103.2541H} pointed out an underoccurrence of hot jupiters at about three days seen by \textit{Kepler} compared with Doppler surveys \citep{2010Sci...330..653H, 2011arXiv1109.2497M}. This study assumed that impostors are negligible. As shown in Fig. \ref{fig1}, our sample, cleaned from impostors, presents a hot jupiter pile-up around three-day periods. The underoccurrence found by \citet{2011arXiv1103.2541H} might be explained by the 35\% of false positives that dilute the hot jupiter pile-up found with Doppler surveys that was confirmed by our cleaned sample.\\

Our measurement of the \textit{Kepler} FPP, significantly higher than previous estimates for close-in giant planets, pointed out that impostors are not negligible in the \textit{Kepler} list of candidates. However, because we measured the FPP only for close-in giant planet candidates, we cannot extrapolate it toward the smaller planets with our data. But we might expected that the FPP for smaller candidates is also higher than the \citetalias{2011ApJ...738..170M} estimation \citep{2012arXiv1207.2481C}. Because there is no reason for the FPP to be constant throughout the entire candidate parameter space (orbital period, transit depth, \textit{Kepler} magnitude, galactic coordinates), exoplanet statistics based on candidates might be biased toward the false-positive distributions.\\  

The latter statement is not true for candidates in multiple systems, for which the FPP is expected to decrease significantly, as pointed out by \citet{2011ApJ...732L..24L}, \citet{2012ApJ...750..112L}, and \citet{2012arXiv1202.6328F}. We note that about one third of the \textit{Kepler} candidates are in multiple systems \citep{2011ApJS..197....8L}.\\

\subsection{Hot-jupiter occurrence rate}
\label{sect:occurrence}
Using our false-positive rate for close-in giant candidates, we tried to estimate the occurrence of hot jupiters in the \textit{Kepler} data, corrected for false positives. We first of all tried to reproduce the \citet{2011arXiv1103.2541H} results for hot jupiters. The authors found an occurrence of $4\pm1$ hot jupiters per thousand GK dwarfs. We assumed the same range of parameters as \citet{2011arXiv1103.2541H} : 4100 K $<$ \teff $<$ 6100 K, 4.0 $<$ \logg $<$ 4.9, period $<$ 10 days, K$_{p} < 15$, 8 \Rearth $<$ R$_{p} <$ 32 \Rearth and no false positive. We corrected the number of candidates detected by their respective $a/R_{\star}$ to take the transit probability into account. We found an occurrence of $5.1\pm0.3$ hot jupiters per thousand stars. This discrepancy with \citet{2011arXiv1103.2541H} might be explained by the improvement in the parameters of \textit{Kepler} candidates and stellar parameters by \citetalias{2012arXiv1202.5852B} compared with \citet{2011ApJ...736...19B}. We note that these two estimates are compatible within 1-$\sigma$. They slightly differ from the $8.9 \pm 3.6$ value found by \citet{2011arXiv1109.2497M} who considered planets with mass greater than 50 M$_{\oplus}$ and a period of less than 11 days. If we considered a FPP of 18.2\% for candidates with orbital periods of less than 10 days, we would find an occurrence of $4.1\pm0.7$ hot jupiters per thousand stars. Now, if we also considered the F-type stars (4100 K $<$ \teff $<$ 7100 K), we would find a lower occurrence of $3.7\pm0.5$ hot jupiters per thousand stars.\\

Considering candidates based on their expected radius may affect the estimate by up to 30\% , due to uncertainties on stellar parameters \citep{2011ApJ...736...19B}. We therefore selected candidates based on their measured transit depth. For that purpose, we selected candidates with a transit depth between 0.4\% and 3\% orbiting around G and K main-sequence stars (4100 K $<$ \teff $<$ 6100 K, 4.0 $<$ \logg $<$ 4.9, K$_{p} < 15$) with periods of less than 10 days. Considering 18.2\%, of impostors we found an occurrence of $6.7\pm0.8$ hot jupiters per thousand GK dwarfs and $5.7\pm0.7$ hot jupiter per thousand FGK dwarfs. Finally, considering candidates with periods of up to 25 days and 35\% of impostors, we found an occurrence of $9.0\pm1.2$ close-in jupiters per thousand GK dwarfs. Occurrences and selection criteria are listed in Table \ref{occurrence}. The occurrence rate of hot jupiters around GK dwarfs when considering candidates from their measured transit depth appears to agree better with radial velocity surveys \citep[see][and reference therein]{2012arXiv1205.2273W} than selecting them according to their estimated radius. We also note that occurrences considering F dwarfs might be diluted by a significant amount of subgiant stars, which are misclassified in the \textit{Kepler} Input Catalog as main-sequence stars, for which giant planets would produce transit depths shallower than 0.4\% \citep[e.g. KOI-428,][]{2011A&A...528A..63S}.\\

\begin{table}[h]
  \centering 
  \caption{Occurrence of hot jupiter per thousand of stars}\label{occurrence}
\renewcommand{\footnoterule}{}                          
\begin{minipage}[c]{\columnwidth} 
\begin{tabular}{ccc}
\hline
 selection criteria  & GK dwarfs$^{\dag}$ & FGK dwarfs$^{\ddag}$ \\
\hline
 8 \Rearth $<$ R$_{p} <$ 32 \Rearth, $P<10d^{\ast}$ & $4.1\pm0.7$ & $3.7\pm0.5$ \\
$0.4\% < \delta < 3\%$, $P<10d^{\ast}$  & $6.7\pm0.8$ & $5.7\pm0.7$ \\
$0.4\% < \delta < 3\%$, $P<25d^{\ast\ast}$  & $9.0\pm1.2$ & $7.9\pm1.0$ \\
\hline
\end{tabular}\\
\footnoterule{$^{\dag}$ 4100 K $<$ \teff $ <$ 6100 K, 4.0 $<$ \logg\ $ <$ 4.9, K$_{p} < 15$\\}
\footnoterule{$^{\ddag}$ 4100 K $<$ \teff $ <$ 7100 K, 4.0 $<$ \logg\ $ <$ 4.9, K$_{p} < 15$\\}
\footnoterule{$^{\ast}$ considering a planet rate of 82.8\%\\}
\footnoterule{$^{\ast\ast}$ considering a planet rate of 65\%}
\end{minipage}
\end{table}

\subsection{Toward smaller candidates}

A similar study on shallower transiting candidates with high-precision instruments such as HARPS-N on the 3.6-m TNG telescope or HiReS on the Keck-1 telescope might constrain the true FPP value of \textit{Kepler} small candidates. As an example, we estimated the HARPS-N radial velocity uncertainty for a 1-hour exposure as a function of stellar magnitude based on our experiment with HARPS for the \textit{CoRoT} follow-up \citep{2011EPJWC..1102001S} and assuming a non-rotating and non-active solar-type star and a systematic error of 50\cms\ at high S/N \citep{2011A&A...534A..58P}. We selected all  \textit{Kepler} candidates with an estimated planetary radius smaller than 5 \Rearth. We assumed for these small planet candidates a density of half the earth density, hence in between Neptune-like and Earth-like planets. We considered that the planet is detected in radial velocity if the amplitude (peak-to-peak) is greater than 3 times the RV uncertainty.\\

Out of the 1981 multi-transit candidates with a radius smaller than 5 \Rearth, we found
that only 77 candidates, hence less than 4\%, may be detected in radial velocity,
including only 26 targets brighter than magnitude 12, and including only 16 candidates
with radius smaller than 3 \Rearth. A more accurate estimation of the FPP than \citetalias{2011ApJ...738..170M} toward the whole population of \textit{Kepler} candidates is thus mandatory for studying the diversity of planets in the \textit{Kepler} field.\\

\section{Conclusion}

From the overall list of 2321 \textit{Kepler} candidates from \citet{2011ApJ...736...19B} and \citetalias{2012arXiv1202.5852B}, we selected 46 that fullfil the criterion of having a transit depth greater than 0.4\%, an orbital period of less than 25 days, a host star brighter than $K_{p} = 14.7$ and a vetting flag different from four. With the SOPHIE spectrograph at the Observatoire de Haute-Provence we observed 28 of them that were not previously announced as planets or followed up by other teams. We found nine new planets \citep[][H\'ebrard et al. in prep.]{bouchy2011, Santerne2011, bonomo2012} which increases the number of secured planets in this sample to 20. We also found two interesting transiting brown dwarfs in the range 25 -- 80 \Mjup (D\'iaz et al. in prep.) which increases the number of objects in this range to five. We also found five undiluted eclipsing binaries and six clearly diluted eclipsing binaries (triple hierarchical systems or background eclipsing binaries). We cannot conclude on a planetary or false-positive scenario for 13 of them due to photon noise limitations or lack of observations. More data with SOPHIE, with HiReS on Keck, or with the new HARPS-N spectrograph mounted on the TNG-3.6m telescope would permit us to conclude on these objets. If we assumed that these 13 candidates have the same proportion of planets and blends as the observed one, we can conclude that the false-positive rate of \textit{Kepler} giant planets on an orbital period of less than 25 days is 34.8\%$\pm$ 6.5\%. This value is clearly incompatible with the FPP $\sim 5 \%$ estimated by \citet{2011ApJ...738..170M} who did not take into account the probability that there might be an undiluted eclipsing binary in the \textit{Kepler} data and they also underestimated the probability of having a diluted binary.\\

Comparing the distribution of planets and binaries found by radial velocity surveys, we estimate the FPP to be quite constant for giant planet candidates with orbital period of less than $\sim 200$ days. Only the RV follow-up of a significant fraction of these long-period candidates can support this statement.\\

We note that if we were to remove the magnitude constraint and the vetting flag constraint from our selection criteria, this sample would increases by 85 new candidates to a total number of 131 giant planet candidates. Following-up these 85 candidates will substantiate our FPP value. This requires a larger telescope, such as the TNG-3.6m telescope with HARPS-N, or the Keck telescope with HiReS to follow these candidates up to magnitude Kp=16.2. This work indicates that only the RV follow-up of a substantial amount of \textit{Kepler} candidates will provide the real value of the FPR.\\

Only a small fraction of \textit{Kepler} small candidates are suited for the radial velocity follow-up. These candidates should be followed in radial velocity to constrain the FPP value for small candidates and to fill the mass-radius diagram of Neptune- and super-Earth like planets. The global FPP values are required for correctly deriving and discussing the distribution of transiting planet parameters.\\

In this paper, we also provided a list of clearly diluted or undiluted binaries. Their analysis can contribute to improve the planet validation techniques \citep[e.g.][]{2011ApJS..197....5F}. \textit{Spitzer} Space Telescope observations of these candidates should reveal a significant depth difference compared with \textit{Kepler} \citep{2012AAS...21941402D}.

\begin{acknowledgements}
We thank the technical team at the Observatoire de Haute-Provence for their support with the SOPHIE instrument and the 1.93-m telescope and in particular for the essential work of the night assistants. We are grateful to the \textit{Kepler} team for giving public access to \textit{Kepler} light curves and for publishing a list of interesting planetary candidates to follow-up. Financial support for the SOPHIE observations from the ÒProgramme National de Plan\'etologieÓ (PNP) of CNRS/INSU, France is gratefully acknowledged. We also acknowledge support from the French National Research Agency (ANR-08-JCJC-0102-01). R.F.D. is supported by CNES. NCS acknowledges the support by the European Research Council/European Community under the FP7 through Starting Grant agreement number 239953, as well as from Funda\c{c}\~ao para a Ci\^encia e a Tecnologia (FCT) through program Ci\^encia\,2007 funded by FCT/MCTES (Portugal) and POPH/FSE (EC), and in the form of grants reference PTDC/CTE-AST/098528/2008 and PTDC/CTE-AST/098604/2008. A. S. thanks Christophe Lovis for fruitful discussions about hot jupiter occurrence. We thank Tim Morton for constructive discussions on M\&J11 results as well as the anonymous referee for his/her comments that improved the quality of this paper.\\

This research has made use of the Exoplanet Data Explorer at exoplanets.org and the NASA Exoplanet Archive, which is operated by the California Institute of Technology, under contract with the National Aeronautics and Space Administration under the Exoplanet Exploration Program. This publication makes use of data products from the Wide-field Infrared Survey Explorer and Two Micron All Sky Survey, which are joint projects of the University of California, Los Angeles, and the Jet Propulsion Laboratory/California Institute of Technology, and respectively the University of Massachusetts and the Infrared Processing and Analysis Center/California Institute of Technology, funded by the National Aeronautics and Space Administration and/or the National Science Foundation.
\end{acknowledgements}

\onltab{2}{
\begin{table*}[h]
\centering
\begin{minipage}[t]{15cm} 
\caption{SOPHIE measurements of KOI-12.}
\begin{tabular}{cccccccc}
\hline
\hline
BJD & RV & $\pm 1\sigma_\mathrm{rv}$ & $V_\mathrm{span}$ & $\pm 1\sigma_{V_\mathrm{span}}$ & Texp & S/N/pix \\
(-2 400 000) & [\kms] & [\kms] & [\kms] & [\kms] & [s] & @550nm \\
\hline
55617.71239  &  -19.132  &  0.799  &  -2.037  &  2.396  &  1006  &  23.1\\
55681.53719  &  -18.278  &  0.427  &  0.079  &  1.282  &  1312  &  43.1\\
\hline
\hline
\label{koi12}
\end{tabular}
\vspace{-0.3cm}
\end{minipage}
\end{table*}}

\onltab{3}{
\begin{table*}[h]
\centering
\begin{minipage}[t]{15cm} 
\caption{SOPHIE measurements of KOI-131.}
\begin{tabular}{ccccccccc}
\hline
\hline
BJD & RV & $\pm 1\sigma_\mathrm{rv}$ & $V_\mathrm{span}$ & $\pm 1\sigma_{V_\mathrm{span}}$ & Texp & S/N/pix \\
(-2 400 000) & [\kms] & [\kms] & [\kms] & [\kms] & [s] & @550nm \\
\hline
56010.59380  &  -9.886  &  0.449  &  0.384  &  1.034  &  1800  &  11.6\\
56012.64402  &  -8.991  &  0.417  &  -0.716  &  0.959  &  1504  &  10.1\\
\hline
\hline
\label{koi131}
\end{tabular}
\vspace{-0.3cm}
\end{minipage}
\end{table*}
}

\onltab{4}{
\begin{table*}[h]
\centering
\begin{minipage}[t]{15cm} 
\caption{SOPHIE measurements of KOI-190.}
\begin{tabular}{cccccccc}
\hline
\hline
BJD & RV$_{A}$ & $\pm 1\sigma_\mathrm{rv_{A}}$ & RV$_{B}$ & $\pm 1\sigma_\mathrm{rv_{B}}$ & Texp & S/N/pix \\
(-2 400 000) & [\kms] & [\kms] & [\kms] & [\kms] & [s] & @550nm \\
\hline
55686.53761  &  -28.810  &  0.220  &  -42.267  &  0.101  &  3600  &  13.8\\
55705.50173  &  -29.612  &  0.161  &  -14.221  &  0.070  &  2525  &  19.0\\
\hline
\hline
\label{koi190}
\end{tabular}
\vspace{-0.3cm}
\end{minipage}
\end{table*}
}

\onltab{5}{
\begin{table*}[h]
\centering
\begin{minipage}[t]{15cm} 
\caption{SOPHIE measurements of KOI-192.}
\begin{tabular}{ccccccccc}
\hline
\hline
BJD & RV & $\pm 1\sigma_\mathrm{rv}$ & $V_\mathrm{span}$ & $\pm 1\sigma_{V_\mathrm{span}}$ & Texp & S/N/pix \\
(-2 400 000) & [\kms] & [\kms] & [\kms] & [\kms] & [s] & @550nm \\
\hline
55754.44878  &  -24.346  &  0.024  &  -0.010  &  0.055  &  2274  &  17.3\\
55770.50760  &  -24.328  &  0.017  &  -0.065  &  0.040  &  3304  &  17.4\\
\hline
\hline
\label{koi192}
\end{tabular}
\vspace{-0.3cm}
\end{minipage}
\end{table*}
}

\onltab{6}{
\begin{table*}[h]
\centering
\begin{minipage}[t]{15cm} 
\caption{SOPHIE measurements of KOI-197.}
\begin{tabular}{ccccccccc}
\hline
\hline
BJD & RV & $\pm 1\sigma_\mathrm{rv}$ & $V_\mathrm{span}$ & $\pm 1\sigma_{V_\mathrm{span}}$ & Texp & S/N/pix \\
(-2 400 000) & [\kms] & [\kms] & [\kms] & [\kms] & [s] & @550nm \\
\hline
55687.53172  &  -15.709  &  0.012  &  -0.003  &  0.028  &  3600  &  21.0\\
55765.49012  &  -15.748  &  0.025  &  -0.025  &  0.057  &  3600  &  15.2\\
55774.47483  &  -15.709  &  0.012  &  0.049  &  0.027  &  3600  &  21.2\\
55801.49572  &  -15.717  &  0.023  &  0.112  &  0.054  &  3600  &  15.5\\
55802.39087  &  -15.749  &  0.010  &  -0.040  &  0.023  &  3600  &  23.4\\
55804.45417  &  -15.725  &  0.015  &  0.003  &  0.034  &  3101  &  17.5\\
55806.39759  &  -15.751  &  0.013  &  0.106  &  0.031  &  3600  &  19.1\\
55810.40145  &  -15.773  &  0.026  &  -0.034  &  0.060  &  3600  &  12.7\\
55828.43357  &  -15.718  &  0.014  &  -0.045  &  0.031  &  3600  &  22.1\\
55831.39958  &  -15.732  &  0.012  &  0.065  &  0.028  &  3600  &  21.0\\
55833.37105  &  -15.746  &  0.014  &  0.040  &  0.033  &  3600  &  18.7\\
55857.35522  &  -15.753  &  0.019  &  0.056  &  0.044  &  3600  &  17.9\\
\hline
\hline
\label{koi197}
\end{tabular}
\vspace{-0.3cm}
\end{minipage}
\end{table*}
}

\onltab{7}{
\begin{table*}[h]
\centering
\begin{minipage}[t]{15cm} 
\caption{SOPHIE measurements of KOI-201.}
\begin{tabular}{ccccccccc}
\hline
\hline
BJD & RV & $\pm 1\sigma_\mathrm{rv}$ & $V_\mathrm{span}$ & $\pm 1\sigma_{V_\mathrm{span}}$ & Texp & S/N/pix \\
(-2 400 000) & [\kms] & [\kms] & [\kms] & [\kms] & [s] & @550nm \\
\hline
55983.68634  &  -60.192  &  0.035  &  0.186  &  0.080  &  1800  &  9.4\\
55989.68601  &  -60.217  &  0.017  &  -0.011  &  0.038  &  1800  &  17.4\\
\hline
\hline
\label{koi201}
\end{tabular}
\vspace{-0.3cm}
\end{minipage}
\end{table*}
}

\onltab{8}{
\begin{table*}[h]
\centering
\begin{minipage}[t]{15cm} 
\caption{SOPHIE measurements of KOI-340.}
\begin{tabular}{cccccccccc}
\hline
\hline
BJD & RV & $\pm 1\sigma_\mathrm{rv}$ & $V_\mathrm{span}$ & $\pm 1\sigma_{V_\mathrm{span}}$ & Texp & S/N/pix \\
(-2 400 000) & [\kms] & [\kms] & [\kms] & [\kms] & [s] & @550nm \\
\hline
55802.54571  &  -99.412  &  0.043  &  -0.089  &  0.098  &  600  &  10.3\\
55857.38486  &  -72.180  &  0.040  &  0.023  &  0.091  &  566  &  10.9\\
\hline
\hline
\label{koi340}
\end{tabular}
\vspace{-0.3cm}
\end{minipage}
\end{table*}
}

\onltab{9}{
\begin{table*}[h]
\centering
\begin{minipage}[t]{15cm} 
\caption{SOPHIE measurements of KOI-418.}
\begin{tabular}{cccccccccc}
\hline
\hline
BJD & RV & $\pm 1\sigma_\mathrm{rv}$ & $V_\mathrm{span}$ & $\pm 1\sigma_{V_\mathrm{span}}$ & Texp & S/N/pix \\
(-2 400 000) & [\kms] & [\kms] & [\kms] & [\kms] & [s] & @550nm \\
\hline
55984.69997  &  -32.135  &  0.077  &  1.426  &  0.177  &  1800  &  7.8\\
55996.68516  &  -32.679  &  0.031  &  -0.035  &  0.071  &  2700  &  17.9\\
\hline
\hline
\label{koi418}
\end{tabular}
\vspace{-0.3cm}
\end{minipage}
\end{table*}
}

\onltab{10}{
\begin{table*}[h]
\centering
\begin{minipage}[t]{15cm} 
\caption{SOPHIE measurements of KOI-419.}
\begin{tabular}{cccccccccc}
\hline
\hline
BJD & RV & $\pm 1\sigma_\mathrm{rv}$ & $V_\mathrm{span}$ & $\pm 1\sigma_{V_\mathrm{span}}$ & Texp & S/N/pix \\
(-2 400 000) & [\kms] & [\kms] & [\kms] & [\kms] & [s] & @550nm \\
\hline
55802.52069  &  -44.153  &  0.042  &  -0.045  &  0.096  &  900  &  10.3\\
55811.35898  &  22.576  &  0.053  &  -0.023  &  0.121  &  900  &  5.4\\
55831.46019  &  22.966  &  0.041  &  0.042  &  0.095  &  900  &  8.0\\
55973.69055  &  6.089  &  0.048  &  0.430  &  0.110  &  1703  &  4.0\\
55977.70619  &  -29.293  &  0.060  &  -0.067  &  0.138  &  900  &  7.1\\
56011.67515  &  31.153  &  0.066  &  0.034  &  0.152  &  900  &  8.0\\
56026.66141  &  -34.940  &  0.050  &  -0.235  &  0.115  &  900  &  5.1\\
\hline
\hline
\label{koi419}
\end{tabular}
\vspace{-0.3cm}
\end{minipage}
\end{table*}
}

\onltab{11}{
\begin{table*}[h]
\centering
\begin{minipage}[t]{15cm} 
\caption{SOPHIE measurements of KOI-425.}
\begin{tabular}{cccccccc}
\hline
\hline
BJD & RV$_{A}$ & $\pm 1\sigma_\mathrm{rv_{A}}$ & RV$_{B}$ & $\pm 1\sigma_\mathrm{rv_{B}}$ & Texp & S/N/pix \\
(-2 400 000) & [\kms] & [\kms] & [\kms] & [\kms] & [s] & @550nm \\
\hline
55802.53448  &  -16.231  &  0.597  &  -29.718  &  0.442  &  900  &  8.4\\
55830.49421  &  -16.402  &  1.061  &  -18.546  &  0.410  &  900  &  6.9\\
\hline
\hline
\label{koi425}
\end{tabular}
\vspace{-0.3cm}
\end{minipage}
\end{table*}
}

\onltab{12}{
\begin{table*}[h]
\centering
\begin{minipage}[t]{15cm} 
\caption{SOPHIE measurements of KOI-607.}
\begin{tabular}{cccccccccc}
\hline
\hline
BJD & RV & $\pm 1\sigma_\mathrm{rv}$ & $V_\mathrm{span}$ & $\pm 1\sigma_{V_\mathrm{span}}$ & Texp & S/N/pix \\
(-2 400 000) & [\kms] & [\kms] & [\kms] & [\kms] & [s] & @550nm \\
\hline
55828.46352  &  -21.323  &  0.029  &  0.073  &  0.067  &  900  &  13.3\\
55830.45215  &  -5.058  &  0.067  &  -0.121  &  0.153  &  900  &  7.5\\
\hline
\hline
\label{koi607}
\end{tabular}
\vspace{-0.3cm}
\end{minipage}
\end{table*}
}

\onltab{13}{
\begin{table*}[h]
\centering
\begin{minipage}[t]{15cm} 
\caption{SOPHIE measurements of KOI-609.}
\begin{tabular}{ccccccc}
\hline
\hline
BJD & RV$_{A}$ & $\pm 1\sigma_\mathrm{rv_{A}}$ & RV$_{B}$ & $\pm 1\sigma_\mathrm{rv_{B}}$ & Texp & S/N/pix \\
(-2 400 000) & [\kms] & [\kms] & [\kms] & [\kms] & [s] & @550nm \\
\hline
55830.43654  &  0.074  &  0.413  &  35.830  &  1.590  &  900  &  7.9\\
55831.44062  &  -0.225  &  0.206  &  18.097  &  0.746  &  900  &  8.3\\
\hline
\hline
\label{koi609}
\end{tabular}
\vspace{-0.3cm}
\end{minipage}
\end{table*}
}

\onltab{14}{
\begin{table*}[h]
\centering
\begin{minipage}[t]{15cm} 
\caption{SOPHIE measurements of KOI-611.}
\begin{tabular}{ccccccccc}
\hline
\hline
BJD & RV & $\pm 1\sigma_\mathrm{rv}$ & $V_\mathrm{span}$ & $\pm 1\sigma_{V_\mathrm{span}}$ & Texp & S/N/pix \\
(-2 400 000) & [\kms] & [\kms] & [\kms] & [\kms] & [s] & @550nm \\
\hline
55828.47808  &  -15.749  &  0.089  &  0.031  &  0.205  &  900  &  10.1\\
55830.48073  &  -15.531  &  0.111  &  0.008  &  0.254  &  900  &  8.9\\
\hline
\hline
\label{koi611}
\end{tabular}
\vspace{-0.3cm}
\end{minipage}
\end{table*}
}

\onltab{15}{
\begin{table*}[h]
\centering
\begin{minipage}[t]{15cm} 
\caption{SOPHIE measurements of KOI-698.}
\begin{tabular}{ccccccccc}
\hline
\hline
BJD & RV & $\pm 1\sigma_\mathrm{rv}$ & $V_\mathrm{span}$ & $\pm 1\sigma_{V_\mathrm{span}}$ & Texp & S/N/pix \\
(-2 400 000) & [\kms] & [\kms] & [\kms] & [\kms] & [s] & @550nm \\
\hline
55832.44715  &  -26.362  &  0.084  &  0.777  &  0.192  &  600  &  6.2\\
55975.71173  &  13.455  &  0.069  &  -0.291  &  0.159  &  600  &  7.3\\
55983.70988  &  -60.514  &  0.080  &  0.703  &  0.183  &  1202  &  4.2\\
\hline
\hline
\label{koi698}
\end{tabular}
\vspace{-0.3cm}
\end{minipage}
\end{table*}
}

\onltab{16}{
\begin{table*}[h]
\centering
\begin{minipage}[t]{15cm} 
\caption{SOPHIE measurements of KOI-1786.}
\begin{tabular}{ccccccccc}
\hline
\hline
BJD & RV & $\pm 1\sigma_\mathrm{rv}$ & $V_\mathrm{span}$ & $\pm 1\sigma_{V_\mathrm{span}}$ & Texp & S/N/pix \\
(-2 400 000) & [\kms] & [\kms] & [\kms] & [\kms] & [s] & @550nm \\
\hline
56013.58505  &  -26.488  &  0.060  &  0.655  &  0.138  &  1226  &  8.3\\
56040.59942  &  -18.745  &  0.177  &  -0.595  &  0.407  &  3600  &  13.3\\
56045.60870  &  -2.950  &  0.081  &  -0.025  &  0.185  &  1289  &  9.8\\
56050.59722  &  9.850  &  0.020  &  0.016  &  0.046  &  3600  &  12.1\\
\hline
\hline
\label{koi1786}
\end{tabular}
\vspace{-0.3cm}
\end{minipage}
\end{table*}
}

\onlfig{7}{
\begin{figure}[H]
\begin{center}
\includegraphics[width=\columnwidth]{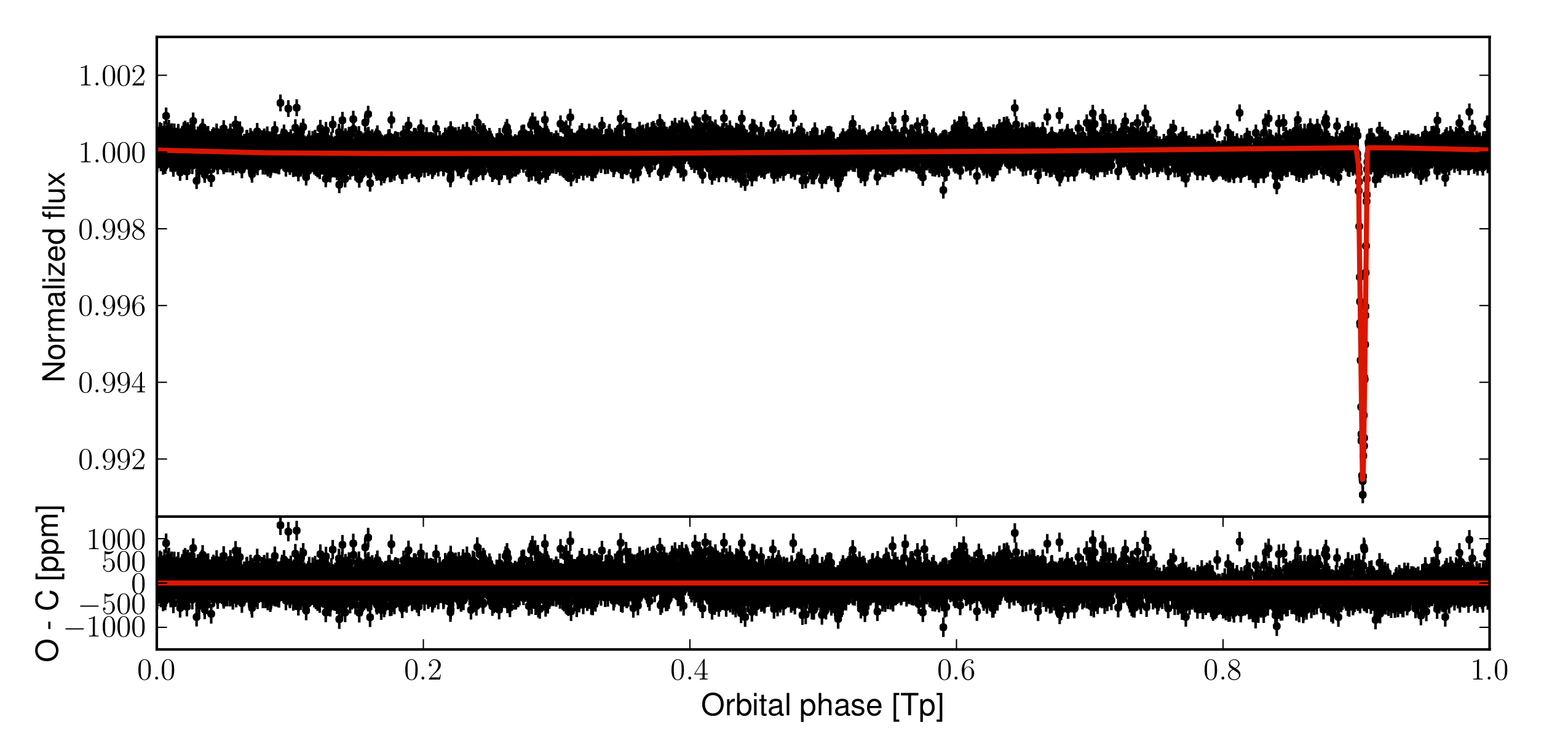}\\
\includegraphics[width=\columnwidth]{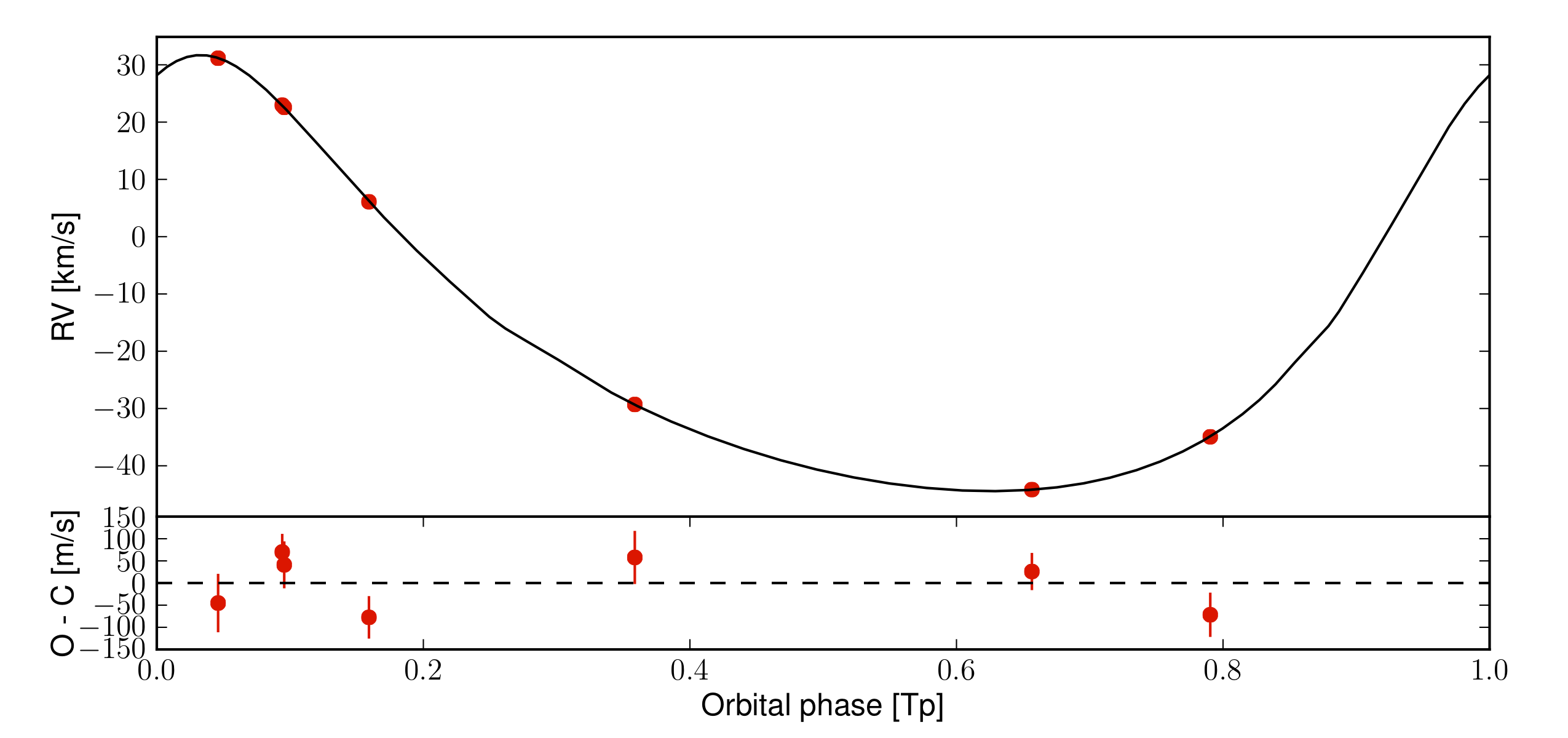}\\
\includegraphics[width=\columnwidth]{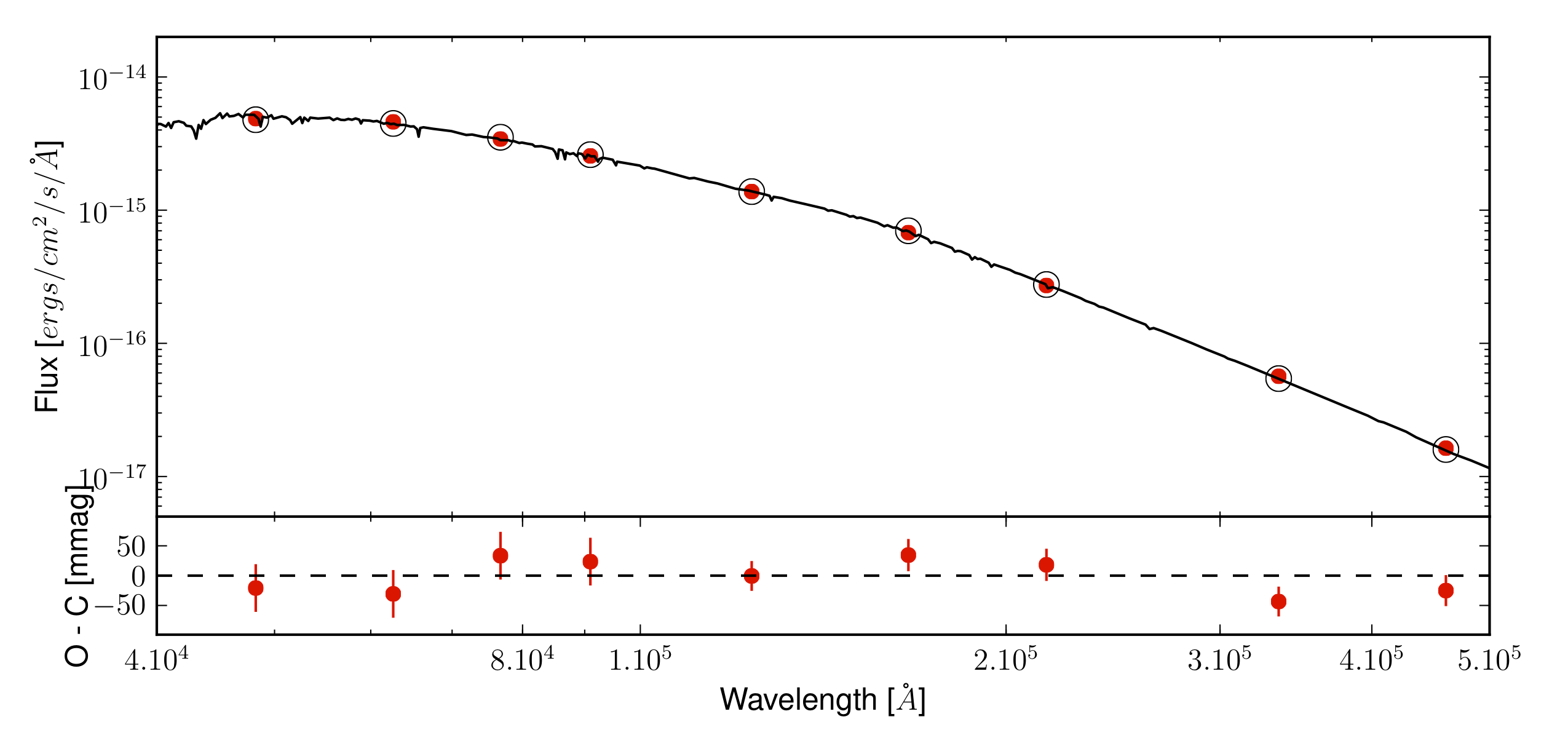}
\caption{KOI-419 light-curve (top panel), radial velocities (middle panel) and spectral energy distribution (bottom panel) with the best-combined fit (red or black line). The phase zero corresponds to the periastron epoch. The open circles in the bottom panel correspond to the best model integrated toward the same band-passes as the photometric measurements (red dots).}
\label{KOI-419}
\end{center}
\end{figure}
}
\vspace{-0.3cm}

\onltab{17}{
\begin{table}[H]
\centering 
\caption{Parameters for KOI-419}
\label{KOI-419params}
\renewcommand{\footnoterule}{}                          
\begin{minipage}{\columnwidth} 
  \begin{tabular}{l c}
\hline
\multicolumn{2}{c}{Primary and secondary stellar parameters} \\
\hline
primary initial mass $M_{i,1}$ [\Msun] &  $1.20 \pm 0.12$$^{\dag}$ \\
secondary initial mass $M_{i,2}$ [\Msun] & $0.70 \pmÊ0.07$$^{\dag}$ \\
$\log(\mathrm{age [yr]})$ &  $ < 8$ \\
metallicity \met [dex]   & 0 (fixed) \\
\hline
\multicolumn{2}{c}{Binary parameters} \\
\hline
distance $d$ [pc] & $1100 \pmÊ15$ \\
reddening $E(B-V)$ & $0.13 \pmÊ0.01$\\
systemic radial velocity $\upsilon_{0}$ [\kms] & $-17.89 \pm 0.03$ \\
\hline
\multicolumn{2}{c}{Orbital parameters} \\
\hline
orbital period $P$ [d] & $20.13151 \pm 0.00001$\\
periastron epoch T$_{p}$ [BJD - 2400000] & $55910.091 \pm 0.008$\\
inclination $i$ [$^{\circ}$] & $87.276 \pm 0.017$\\
eccentricity $e$ & $0.3338 \pm 0.0006$ \\
argument of periastron $\omega$ [$^{\circ}$] & $335.16 \pm 0.16$\\
\hline
\end{tabular}  
\vspace{-0.3cm}
\footnotetext{$^{\dag}$Assuming 10\% uncertainty on the evolution tracks.}
\end{minipage}
\end{table}
}

\onlfig{8}{
\begin{figure}[H]
\begin{center}
\includegraphics[width=\columnwidth]{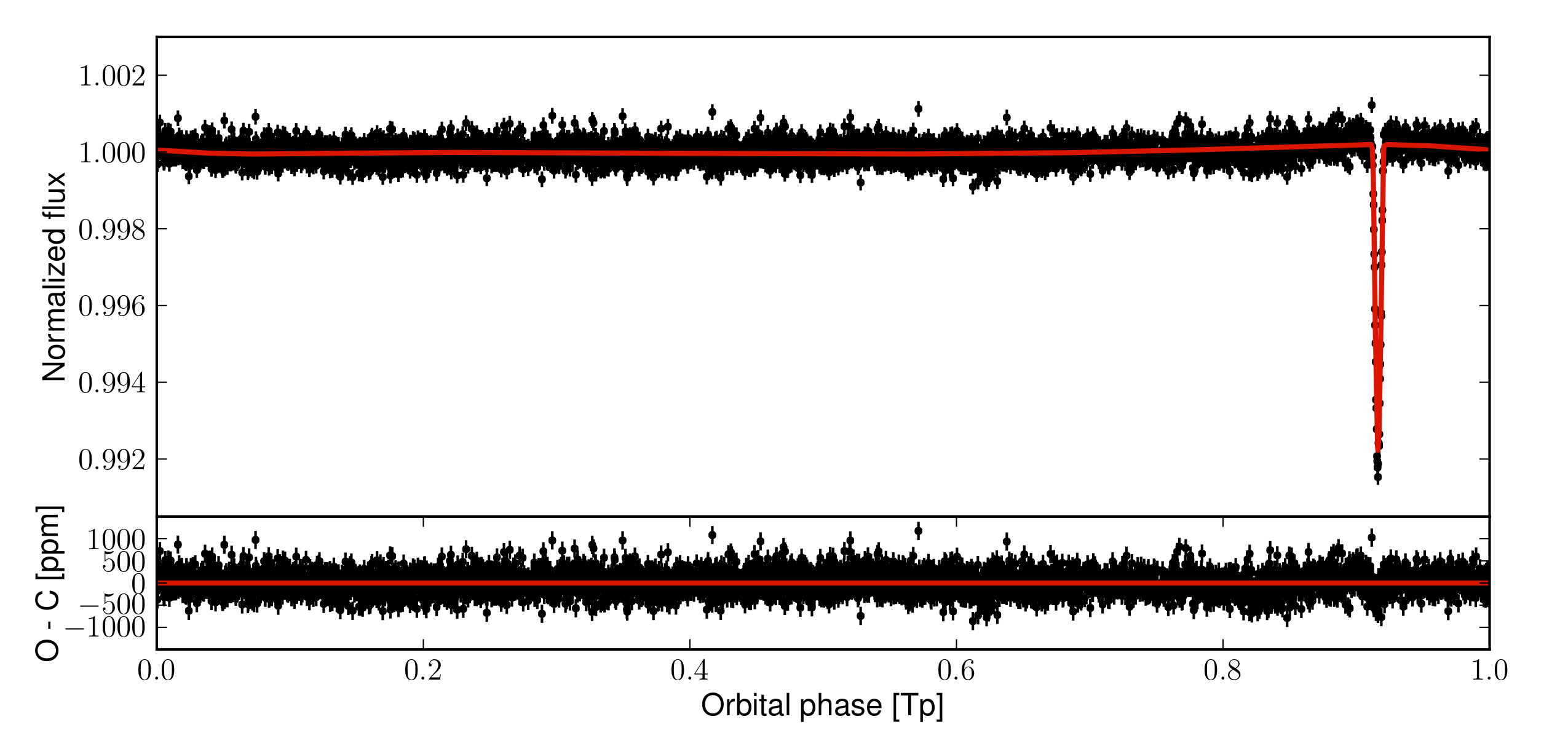}\\
\includegraphics[width=\columnwidth]{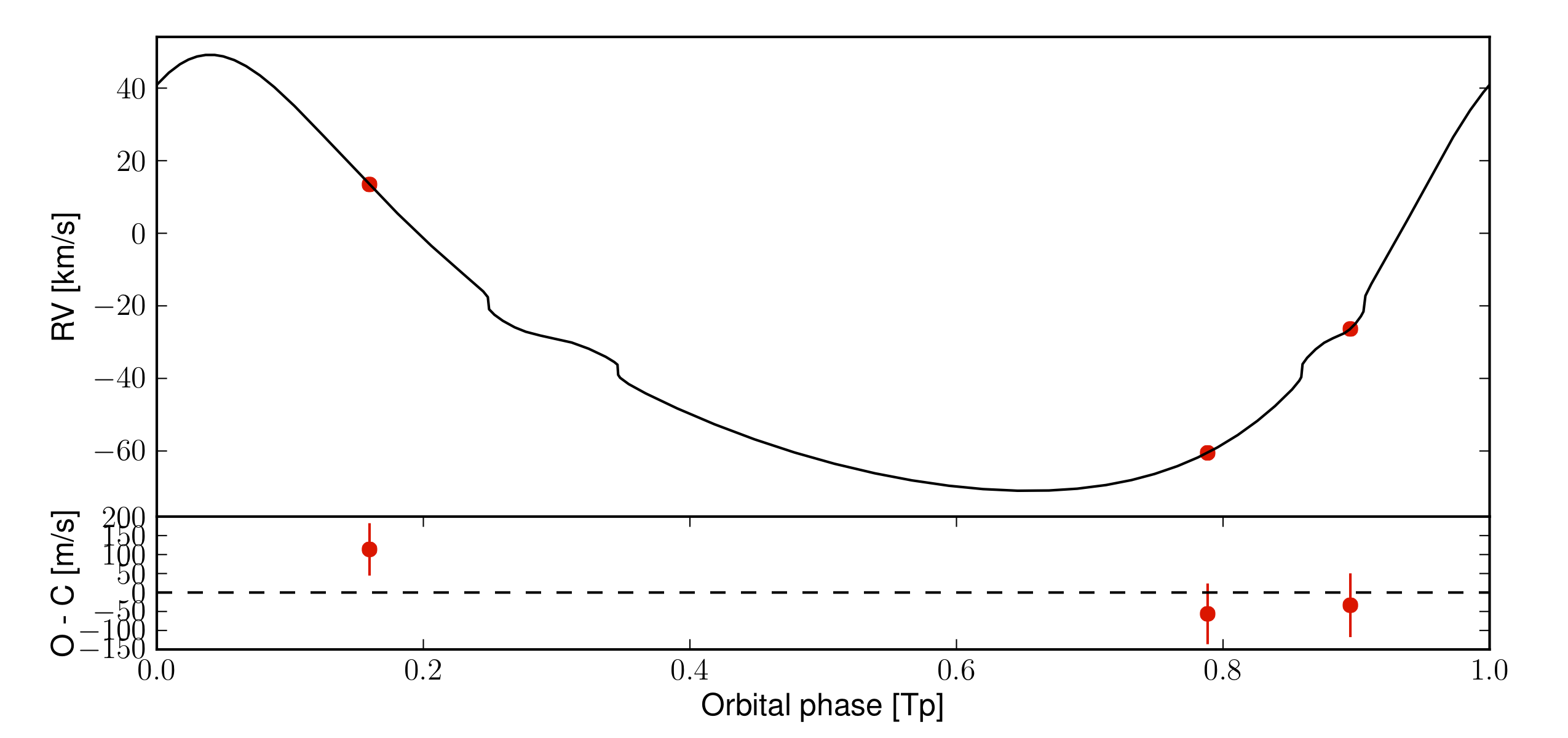}\\
\includegraphics[width=\columnwidth]{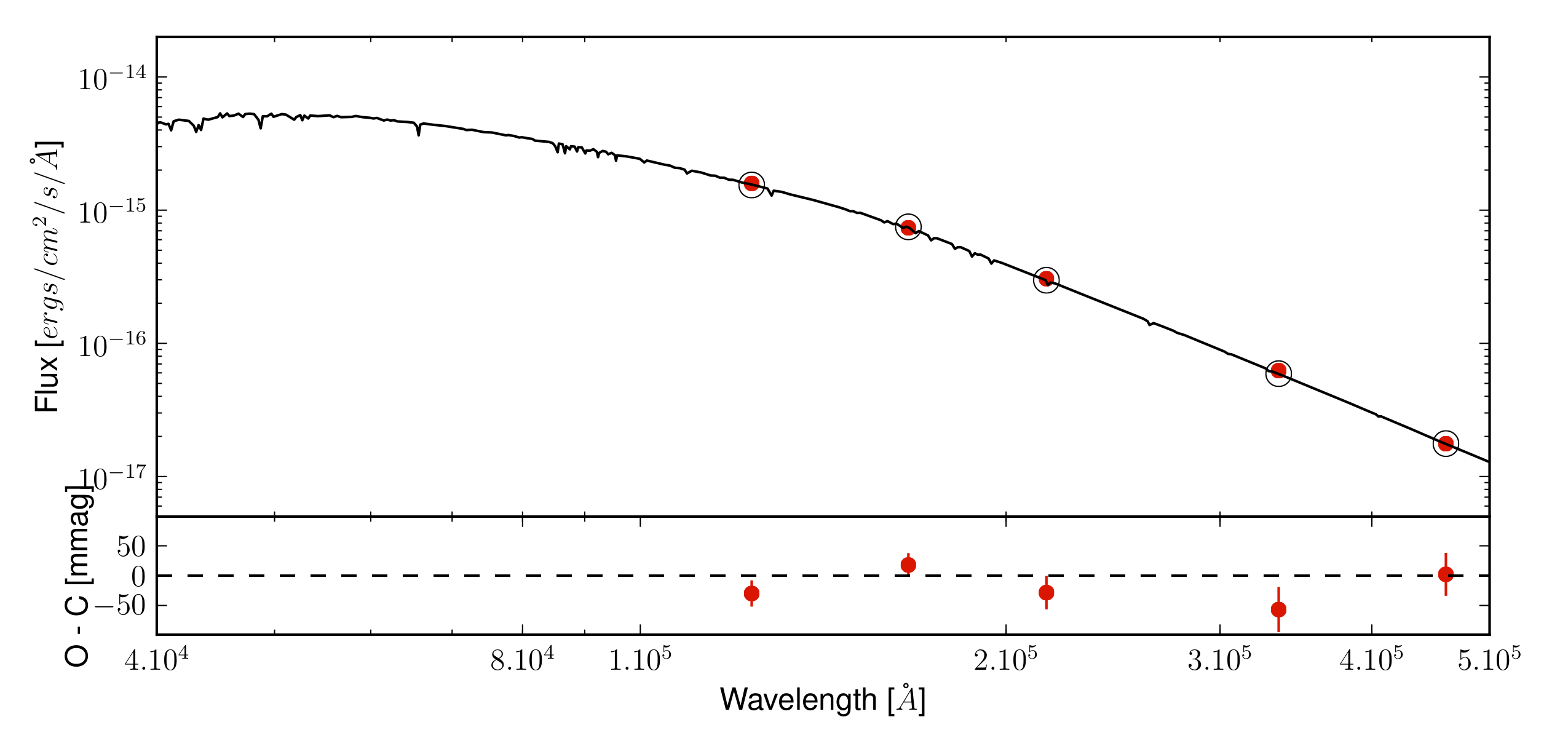}
\caption{KOI-698 light-curve (top panel), radial velocities (middle panel) and spectral energy distribution (bottom panel) with the best-combined fit (red or black line).The phase zero corresponds to the periastron epoch. The open circles in the bottom panel correspond to the best model integrated toward the same band-passes as the photometric measurements (red dots). The Rossiter-McLaughlin-like effect seen around phase 0.3 and 0.9 is due to the expected variation when the primary and secondary spectra are blended.}
\label{KOI-698}
\end{center}
\end{figure}
}
\vspace{-0.3cm}

\onltab{18}{
\begin{table}[H]
\centering 
\caption{Parameters for KOI-698}
\label{KOI-698params}
\renewcommand{\footnoterule}{}                          
\begin{minipage}{\columnwidth} 
  \begin{tabular}{l c}
\hline
\multicolumn{2}{c}{Primary and secondary stellar parameters} \\
\hline
primary initial mass $M_{i,1}$ [\Msun] &  $1.34 \pm 0.13$$^{\dag}$ \\
secondary initial mass $M_{i,2}$ [\Msun] & $1.1 \pmÊ0.1$$^{\dag}$ \\
$\log(\mathrm{age [yr]})$ &  $8.85\pm0.15$$^{\dag}$ \\
metallicity \met [dex]   & 0 (fixed) \\
\hline
\multicolumn{2}{c}{Binary parameters} \\
\hline
distance $d$ [pc] & $1496 \pmÊ32$ \\
reddening $E(B-V)$ & $0.24 \pmÊ0.05$\\
systemic radial velocity $\upsilon_{0}$ [\kms] & $-28.55 \pm 0.13$ \\
\hline
\multicolumn{2}{c}{Orbital parameters} \\
\hline
orbital period $P$ [d] & $12.7187 \pm 0.00001$ \\
periastron epoch T$_{p}$ [BJD - 2400000] & $55045.215 \pm 0.014$\\
inclination $i$ [$^{\circ}$] & $84.43 \pm 0.06$\\
eccentricity $e$ & $0.34 \pm 0.002$ \\
argument of periastron $\omega$ [$^{\circ}$] & $329.54 \pm 0.75$\\
\hline
\end{tabular}  
\vspace{-0.3cm}
\footnotetext{$^{\dag}$Assuming 10\% uncertainty on the evolution tracks.}
\end{minipage}
\end{table}
}


\end{document}